\documentclass[aps,prc,twocolumn,preprintnumbers,showpacs,amsmath,amssymb,superscriptaddress]{revtex4-2}
\usepackage{graphicx}
\usepackage{dcolumn}
\usepackage{bm}
\usepackage{color}
\usepackage{enumitem}

\begin{document}

\title{High voltage and electrode system for a cryogenic experiment
  to search for the neutron electric dipole moment }

\author{M.~A.~Blatnik} 
\affiliation{Los Alamos National Laboratory, Los Alamos, New Mexico 87545, USA}


\author{S.~M.~Clayton} 
\affiliation{Los Alamos National Laboratory, Los Alamos, New Mexico 87545, USA}

\author{S.~A.~Currie} 
\affiliation{Los Alamos National Laboratory, Los Alamos, New Mexico 87545, USA}

\author{B.~W.~Filippone} 
\affiliation{W.~K.~Kellogg Radiation Laboratory, California Institute of Technology, Pasadena, California 91125, USA}

\author{M.~Makela} 
\affiliation{Los Alamos National Laboratory, Los Alamos, New Mexico 87545, USA}

\author{C.~M.~O'Shaughnessy} 
\affiliation{Los Alamos National Laboratory, Los Alamos, New Mexico 87545, USA}

\author{N.~S.~Phan} 
\affiliation{Los Alamos National Laboratory, Los Alamos, New Mexico 87545, USA}

\author{J.~C.~Ramsey} 
\affiliation{Oak Ridge National Laboratory, Oak Ridge, Tennessee 37831, USA}

\author{G.~V.~Riley} 
\affiliation{Los Alamos National Laboratory, Los Alamos, New Mexico 87545, USA}

\author{A.~Roberts} 
\affiliation{Los Alamos National Laboratory, Los Alamos, New Mexico 87545, USA}

\author{T.~Sandborn} 
\affiliation{Los Alamos National Laboratory, Los Alamos, New Mexico 87545, USA}

\author{T.~J.~Schaub} 
\affiliation{Los Alamos National Laboratory, Los Alamos, New Mexico 87545, USA}

\author{G.~M.~Seidel} 
\affiliation{Department of Physics, Brown University, Providence, Rhode Island 02912, USA}

\author{E. Smith} 
\affiliation{Los Alamos National Laboratory, Los Alamos, New Mexico 87545, USA}

\author{I.~L.~Smythe} 
\affiliation{Los Alamos National Laboratory, Los Alamos, New Mexico 87545, USA}

\author{J.~Surbrook} 
\affiliation{Los Alamos National Laboratory, Los Alamos, New Mexico 87545, USA}


\author{W.~Wei} 
\affiliation{Los Alamos National Laboratory, Los Alamos, New Mexico 87545, USA}

\author{W.~Yao} 
\affiliation{Oak Ridge National Laboratory, Oak Ridge, Tennessee 37831, USA}

\author{T.~M.~Ito} 
\email[Corresponding author. Electronic address: ]{ito@lanl.gov}
\affiliation{Los Alamos National Laboratory, Los Alamos, New Mexico 87545, USA}

\date{\today}

\begin{abstract}
The cryogenic approach to the search for the neutron electric dipole moment---performing the experiment in superfluid liquid helium---holds promise for a substantial increase in sensitivity, potentially enabling a sensitivity level of $10^{-28}$~$e\cdot$cm. A crucial component in realizing such an experiment is the high voltage and electrode system capable of providing an electric field of 75~kV/cm. This, in turn, requires an electric potential of 635~kV to be applied to the high voltage electrode, while simultaneously  satisfying other experimental constraints, such as those on heat load and magnetic noise requirements. This paper describes the outcome of a comprehensive development program addressing these challenges. 
It outlines the system requirements, discusses new insights into relevant physical phenomena, and details selected technical solutions with
their corresponding experimental demonstrations and expected performance. The results collectively demonstrate the successful development of the
necessary technology for the high-voltage and electrode system for this approach.
\end{abstract}

\preprint{LA-UR-25-31563}

\maketitle

\section{Introduction}
The existence of a non-zero permanent electric dipole moment (EDM) in a non-degenerate state of a system with spin $J\ne0$ is a violation of
both time-reversal invariance and parity invariance. Based on the $CPT$ theorem, the violation of time-reversal invariance directly implies a violation of
$CP$ invariance, which is the combined operation of charge conjugation and parity.

Searches for new sources of $CP$ violation through searches of EDMs are
strongly motivated for the following reasons (see, e.g., Ref.~\cite{Chupp2019} and references therein): (1)~$CP$ violation is one
of the three essential ingredients for dynamically generating the
observed matter-antimatter asymmetry in the Universe~\cite{Sakharov1991}. However, the
amount of $CP$ violation contained in the Standard Model (SM) of
particle physics through the complex phase of the
Cabibbo-Kobayashi-Maskawa (CKM) matrix is known to be insufficient. Therefore, it is natural to expect additional sources of
$CP$ violation.  (2)~The other source of $CP$ violation contained in
the SM, the so-called $\theta$ term in the quantum chromodynamics
(QCD) Lagrangian, is constrained by nonobservation of the neutron EDM
(nEDM) to be unnaturally small.  (3)~The SM value of EDMs based on the
CKM phase are highly suppressed because in the SM the $CP$ violation
processes are quark flavor-changing at the tree level. Therefore, EDMs
are a sensitive probe of new sources of CP violation.  (4)~Many
extensions of the SM contain new sources of $CP$ violation, predicting
much larger values of EDMs.

The search for neutron EDM (nEDM) was pioneered by Smith, Purcell,
and Ramsey with an experiment performed in 1949~\cite{Smith1957}. In
the subsequent 70 odd years, many more experiments have been performed
employing increasingly refined experimental methods and resulting in 
correspondingly improved sensitivities and upper limits. 
Many of these experiments have searched
for the nEDM by subjecting spin polarized neutrons to uniform static magnetic
and electric fields and looking for a possible change in the spin
precession frequency corresponding to the change in the relative
orientation of the electric field with respect to the magnetic
field. While early experiments were performed using a beam of cold
neutrons, stored ultracold neutrons
(UCNs)~\cite{Ignatovich1990,Golub1991} are used in all the 
recently performed~\cite{Baker2007,Serebrov2015,Abel2020} experiments as well as most of the experiments currently being developed~\cite{Ito2018,Wurm2019,Martin2020,n2EDM_apparatus_2021} to
suppress the effect of the motional magnetic
field~\cite{Lamoreaux2009} (see, however, Ref.~\cite{Piegsa2013}).

The statistical sensitivity of such measurements for a batch of stored UCNs, $\delta d_n$, where $d_n$ stands for nEDM, depends on three quantities, namely, $E$ the strength of the
electric field, $T$ the free precession time, and $N$ the number of
neutrons in the batch, and is expressed as follows:
\begin{equation}
\label{eq:sensitivity}
\delta d_n \propto
  \frac{1}{ET\sqrt{N}}.
\end{equation}
Typically, many such measurements are repeated over the duration of an
experiment. The current upper limit, given by an experiment performed
at Paul Scherrer Institute (PSI nEDM experiment) that produced a
result of $d_n = (0.0 \pm 1.1_{\rm stat} \pm 0.2_{\rm sys})\times
10^{-26}$~$e\cdot$cm, is $|d_n | < 1.8\times 10^{-26}$~$e\cdot$cm (90\%
C.L.)~\cite{Abel2020}. In this experiment, typical values for $E$,
$T$, and $N$ were $E=11$~kV/cm, $T=180$~s, and $N=11,400$.

However, in room-temperature experiments utilizing stored UCNs, the sensitivity of the searches is tightly constrained by the practically achievable
values of those parameters. The strength of the electric field $E$ is limited by electrical breakdown at the electrode-insulator (triple) junction, a constraint where significant improvements in room-temperature setups are not expected. The storage time $T$ is restricted by UCN losses during wall collisions, primarily caused by the upscattering of neutrons due to surface hydrogen contamination. Finally, the number of neutrons $N$ is limited by the performance of
current UCN sources.
A comparison between the PSI nEDM experiment~\cite{Abel2020} and the previous Institute Laue-Langevin (ILL)-Sussex experiment~\cite{Baker2007}, which established the prior upper
limit, shows the limitations mentioned here. The PSI collaboration utilized a modified version of the original ILL-Sussex apparatus at the PSI UCN
source. While the electric field $E$ remained nearly unchanged (11~kV/cm vs 10~kV/cm), the PSI experiment achieved an overall enhancement in the $T\sqrt{N}$ by a factor of 1.25. This gain was primarily a result of a longer storage time (180~s vs 130~s), enabled by the softer UCN spectrum produced by
the PSI source, thus resulting in only a marginal improvement over the final result from the ILL-Sussex experiment of $d_n = (-0.21 \pm 1.82)\times 10^{-26}$~$e\cdot$cm~\cite{Pendlebury2015}.

In order to significantly improve all of $E$, $T$, and $N$, a completely new
method for performing an nEDM experiment was proposed by Golub and
Lamoreaux in 1994~\cite{Golub1994}. In this method, the experiment is
performed in a bath of liquid helium (LHe) at approximately
0.4~K. The unique features of this approach include:
\begin{enumerate}
\item In-situ production of UCNs inside the measurement cells
from a cold neutron beam of 0.89~nm wavelength using the superthermal
process in superfluid liquid helium~\cite{Golub1977}.
\item Use of spin-polarized $^3$He atoms as a comagnetometer.
\item Use of the spin-dependent neutron capture reaction on a $^3$He
  atom ($n+^3{\rm He}\to p+t$) and the resulting LHe scintillation (see,
  e.g., Refs.~\cite{Ito2012} and \cite{Phan2020}) as the analyzer of the neutron
  spin. Note that the cross section for spins antiparallel ($\sigma_{ \uparrow\mathrel{\mspace{-1mu}}\downarrow}$) is much larger than that for spins parallel ($\sigma_{\uparrow\uparrow}$), that is, $\sigma_{ \uparrow\mathrel{\mspace{-1mu}}\downarrow} \gg \sigma_{\uparrow\uparrow}$.
\end{enumerate}
A larger $E$ was expected based on the expectation that LHe would be a better insulator than a vacuum. Additionally, it was hypothesized that LHe would
suppress electrical breakdown processes at the electrode-insulator interface, which had previously limited the electric field strength in room-temperature
nEDM experiments.
Performing a measurement at
cryogenic temperatures suppresses some of the loss mechanisms for
stored UCNs, resulting in longer
$T$~\cite{Ageron1985,Korobkina2004}. Producing UCNs directly in the
experiment from a 0.89~nm cold neutron beam could eliminate UCN loss
due to transport. As long as a sufficiently strong cold neutron beam
is used, a larger $N$ can be expected.

This experiment employs two complementary methods to extract the possible nEDM signal, the free spin precession method and the critical spin dressing method~\cite{Golub1994}. In the free precession method, both neutrons and $^3$He atoms precess freely. The precession frequency of $^3$He atoms is measured by superconducting quantum interference device (SQUID)-based magnetometers. The neutron precession frequency is determined by the combination of the $^3$He precession frequency and the frequency at which LHe scintillation signal (see above) is measured, which indicates the precession frequency difference between the $^3$He atoms and the neutrons. If the application of an electric field changes the relative precession frequency between the two species while no change is observed in the $^3$He precession frequency, it indicates a nonzero nEDM, as the EDM of $^3$He atoms is expected to be much smaller compared to nEDM~\cite{Dzuba2007,Stetcu2008}. On the other hand, in the critical spin dressing method, a strong non-resonant oscillating magnetic field is applied in the direction perpendicular to the holding magnetic field. This modifies the effective gyromagnetic ratios~\cite{Cohen-Tannoudji1969} of both species so that they both precess at the same precession frequency in the absence of an electric field. If the application of an electric field changes their relative precession frequency, signaled by the change in the LHe scintillation signal, this indicates a nonzero nEDM. Using these two complementary methods to extract the nEDM signal could potentially uncover previously unknown systematic effects. 

Following the publication of Ref.~\cite{Golub1994}, a research and
development (R\&D) program was launched with a goal of designing an
experiment based on this method (referred to hereafter as the Cryogenic nEDM
experiment). The R\&D program included physics measurements needed to
validate the concept and to design the experiment, as well as 
efforts to develop various technologies required by the
experiment. The status of the development, as well as the design, as
of 2019 is described in Ref.~\cite{Ahmed2019}. A statistical sensitivity of $(2-3)\times 10^{-28}$~$e\cdot$cm was projected. 

One key component of this R\&D program was the development of the
system that is capable of providing the necessary high electric field to the volume in which UCNs are stored. 
The statistical sensitivity of an EDM experiment is directly proportional to the applied electric field, making its maximization a critical objective. However, the
potential for a significant increase in $E$ by conducting the experiment in LHe had not been experimentally validated at the time of the conception of this new technique. Furthermore, the fundamental mechanisms of electrical breakdown in LHe were not well understood. Consequently, the R\&D  program for the Cryogenic nEDM experiment included a dedicated effort to investigate various aspects of electrical breakdown in LHe to inform the design of
the high-voltage and electrode systems.

The purpose of this paper is to present a comprehensive report on the outcomes of the HV and electrode development program. It includes the results of 
studies of relevant physics phenomena, the system requirements,
the chosen technical solutions, their experimental demonstrations, the
final system design, and the expected performance. Although the
funding for developing the Cryogenic nEDM experiment has been
discontinued, the results presented in this paper are relevant to and
important for other similar efforts---efforts to develop an
nEDM experiment to be performed in LHe, including the one considered
for the European Spallation Source~\cite{Degenkolb2022,Abele2023} and any future experiments that require low magnetic Johnson noise high voltage electrodes in cryogenic environments.

The remainder of this paper is organized as follows. In
Sec.~\ref{sec:requirements}, the requirements for the high-voltage
system and electrode system will be
described. Section~\ref{sec:physical_phenomena} will discuss relevant
physical phenomena, including aspects of electrical breakdown in LHe
and effects of cosmic rays and other ionizing radiation. In
Sec.~\ref{sec:design}, the design of the system will be described along with the performance demonstrated at the component
level. The expected performance of the full system will be discussed
in Sec.~\ref{sec:expected_performance}. In Sec.~\ref{sec:discussion},
conclusion and outlook will be given. 

\section{High voltage and electrode system requirements}
\label{sec:requirements}
\subsection{Overview}
\begin{figure}
    \centering
    \includegraphics[width=1.0\linewidth]{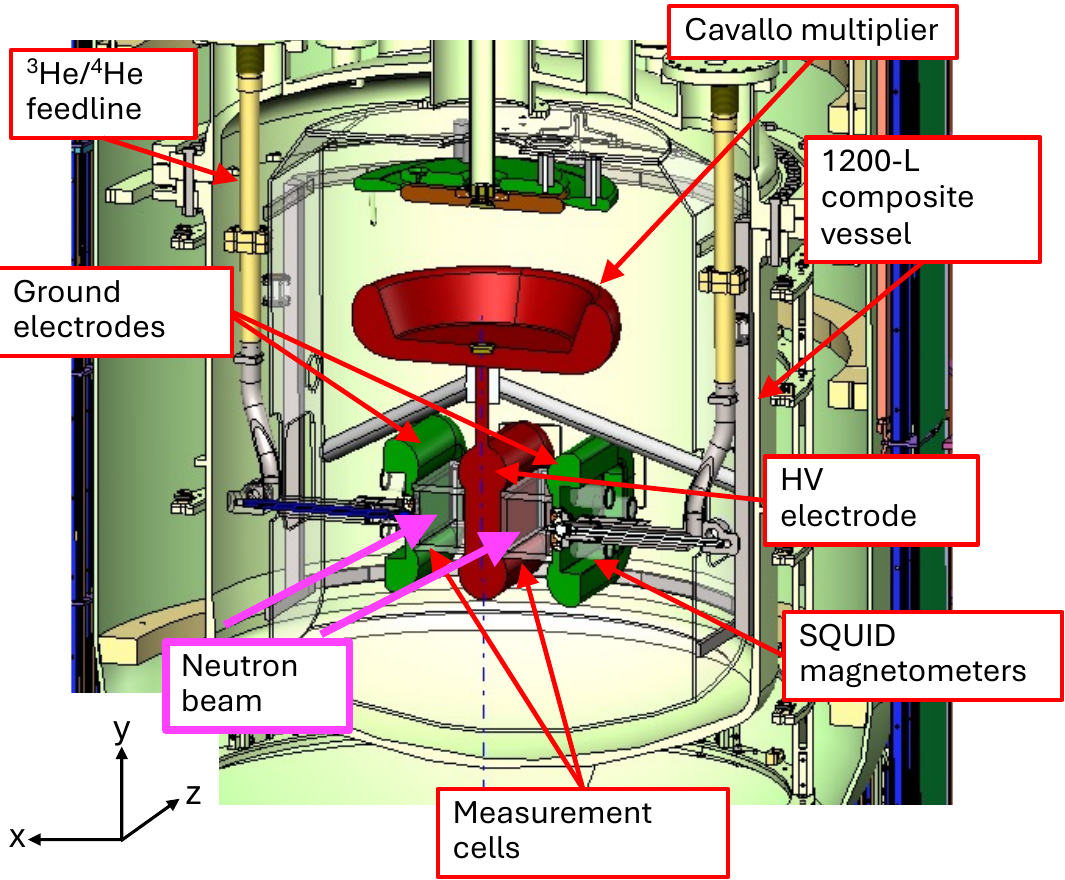}
    \caption{A detailed view of the components of the central detector system within the composite central volume.}
    \label{fig:CDS_schematic}
\end{figure}

The design goal of the high voltage and electrode system was to provide
a stable DC electric field of $E=75$~kV/cm to the region inside the measurement cells, the
volume filled with 0.4~K LHe that stores UCNs~\cite{Ahmed2019}. 
The design called for the following: (1)~the measurement cell walls made of poly(methyl methacrylate) (PMMA) to serve as part of the light collection
system, with dimensions of 10.16~cm (along the electric field)$\times$ 12.70~cm $\times$ 42~cm (along the neutron beam) in outer
dimension with a wall thickness of 1.27~cm, (2)~the two measurement cells sandwiched between electrodes that are roughly $10$~cm~$\times$~30~cm$\times$~60~cm in size, and (3)~the electrodes and measurement cells immersed in 0.4~K LHe. 
These components comprise the Central Detector System (CDS) of the experiment and are shown in Fig.~\ref{fig:CDS_schematic}. Following the convention of the Cryogenic nEDM experiment, we adopt a coordinate system in which the positive z direction points in the direction in which the cold neutron beam travels and the positive y points upwards. The electric and magnetic fields are applied in the $x$ direction. 

Beyond the requirements necessary to achieve a stable electric field of 75 kV/cm, additional constraints on the electrode materials arise from other aspects of the experiment. They are:
\begin{enumerate}
\item The material cannot have too high an electrical conductivity. This requirement is driven, in turn, by two different requirements: (1) the magnetic Johnson noise on the SQUID-based magnetometer to measure the precession frequency of
 spin polarized $^3$He atoms has to be such that it does not degrade the precision with which the precession frequency is determined, and (2) the Joule
 heating from eddy currents due to the radio frequency (RF) field for 
the  dressed spin measurement~\cite{Golub1994} has to be within the allowed heat budget. Specifically, the resistivity of the electrode material needs to be such that the magnetic Johnson noise for the SQUID gradiometer is less than 1~fT/$\sqrt{\rm Hz}$ and the Joule heating from the spin dressing field is less than 6~mW~\cite{Ahmed2019}. Details of how these requirements translate to the resistivity of the material will be discussed in Sec.~\ref{sec:resistivity_requirements}.
\item The material should be non-magnetic. The static magnetic field
 in the region inside the measurement cells, which is approximately
  3~$\mu$T, needs to be uniform to $5\times 10^{-4}$ with
 field gradients smaller than 0.5~nT/m in the direction of the
 static field and 1~nT/m in the direction perpendicular to the
  static field~\cite{Ahmed2019}. Because of this stringent requirement, many of the  so-called ``non-magnetic'' technical materials, such as stainless
 steel and inconel, are disallowed. Also, materials that become
 superconducting cannot be used in general because the field expelled due to the
 Meissner effect would disturb the field uniformity inside the
 measurement cells.
\item 
The material should not exhibit significant neutron absorption that produces particle emissions on the time scale of the measurement. While most of the neutron beam passes through the measurement cells, entering and exiting via the entrance and exit windows, a small fraction is scattered by various components in the beam path, and some of this scattered flux interacts with the electrodes. Such materials would become radioactively activated when the apparatus is exposed to a high-flux neutron beam and subsequently become sources of background radiation.
 
\end{enumerate}


For the geometry considered for the Cryogenic nEDM experiment, an electric field of 75~kV/cm requires an electric potential difference of 635~kV applied between the HV and ground electrodes. This 635~kV could, in principle, be brought in from outside the cryostat using feedthroughs (direct feed method), or it could be generated inside the 0.4-K LHe volume in which the measurement cells are immersed. However, the chosen method needs to meet the same heat load and non-magnetism requirements mentioned above for the electrode materials. In addition, it needs to be compatible with the operation of SQUIDs. These considerations render the possibility of using the direct feed approach impractical and necessitate the development of an in-situ HV generation method based on the concept of an electrostatic induction machine, known as  Cavallo's multiplier~\cite{Clayton2018}. Our implementation of this method for generating 635~kV in the LHe volume is discussed in Sec.~\ref{sec:design}.

\subsection{Electrode resistivity requirements\label{sec:resistivity_requirements}}

\subsubsection{Magnetic Johnson noise}
Magnetic Johnson noise in the context of precision measurements have been investigated by a number of research groups (see, e.g., Refs.~\cite{Varpula1984,Nenonen1996,Henkel2005,Lamoreaux1999,Munger2005,Lee2008,Ayres2021}). These calculations are typically performed using one of two methods: the direct method or reciprocal method. In the direct method, the fluctuating current density is given by the conductivity of the material through the Johnson-Nyquist theorem~\cite{Johnson1928,Nyquist1928}, and the resulting fluctuation magnetic field is calculated either by solving Maxwell's equations~\cite{Varpula1984,Nenonen1996} or by applying the Biot-Savart law~\cite{Lamoreaux1999,Sandin2010,Ready2021}. The reciprocal method, on the other hand, makes use of the fluctuation-dissipation (F-D) theorem~\cite{Callen1951,Kubo1966}, of which the Johnson-Nyquist theorem is a special case. With this method, the magnetic Johnson noise (fluctuation) is determined from the power dissipated in the conductor body (dissipation) when it is driven by an oscillating magnetic field.

Both methods have been used to derive analytic formulas or general expressions for some specific geometries~\cite{Varpula1984,Nenonen1996,Kasai1993,Roth1998,Clem1987,Sidles2003,Lee2008}. However, calculating the magnetic Johnson noise using the direct method for a general case is a daunting task. Since this Cryogenic nEDM experiment has a rather complex electrode geometry (see Fig.~\ref{fig:CDS_schematic}) it is essential in establishing a reliable resistivity requirements that we have a method that allows us to calculate the magnetic Johnson noise in a straightforward manner for arbitrary geometries for a given electrode resistivity. 

In Ref.~\cite{Phan2024} we demonstrated that the reciprocal method, combined with an finite element analysis tool such as COMSOL~\cite{COMSOL}, provides a practical way to calculate magnetic Johnson noise for arbitrary geometries and frequencies. We showed furthermore that this method can be used, not only to calculate magnetic Johnson noise evaluated at a point in space, but also to calculate magnetic Johnson noise detected by a magnetic field detector with a more general shape, such as a finite-sized loop, a gradiometer, etc., taking into account the correlations of magnetic fields detected by different parts of such a detector.

More concretely, when evaluating the magnetic Johnson noise at a given observation point near the conductor, a hypothetical current loop with an oscillating current $I(t)=I_0 \sin \omega t$ and area $A$ is placed at that location. The shape of the current loop corresponds to the shape of the magnetic field detector: a small loop for evaluating the magnetic Johnson noise at a point in space,  a finite-sized loop for evaluating the noise  detected by such a loop, and two loops wound in the opposite direction to evaluate the noise detected by a gradiometer, etc. The resulting oscillating magnetic field incurs a time-averaged power dissipation $P$ in the conductor due to eddy currents. The F-D theorem shows that the magnetic Johnson noise $B_n(f)$ is related to the power dissipation as follows:
\begin{equation}
\label{eq:F-Dtheorem}
    B_n(f) = \frac{\sqrt{4kT}\sqrt{2P(f)}}{2\pi fANI},
\end{equation}
where $T$ is the temperature of the conductor, $k$ is the Boltzmann constant, $N$ is the number of turns in the current loop, and $\omega = 2\pi f$ is the angular frequency of the driving current. 

\begin{figure}
    \centering
    \includegraphics[width=1.0\linewidth]{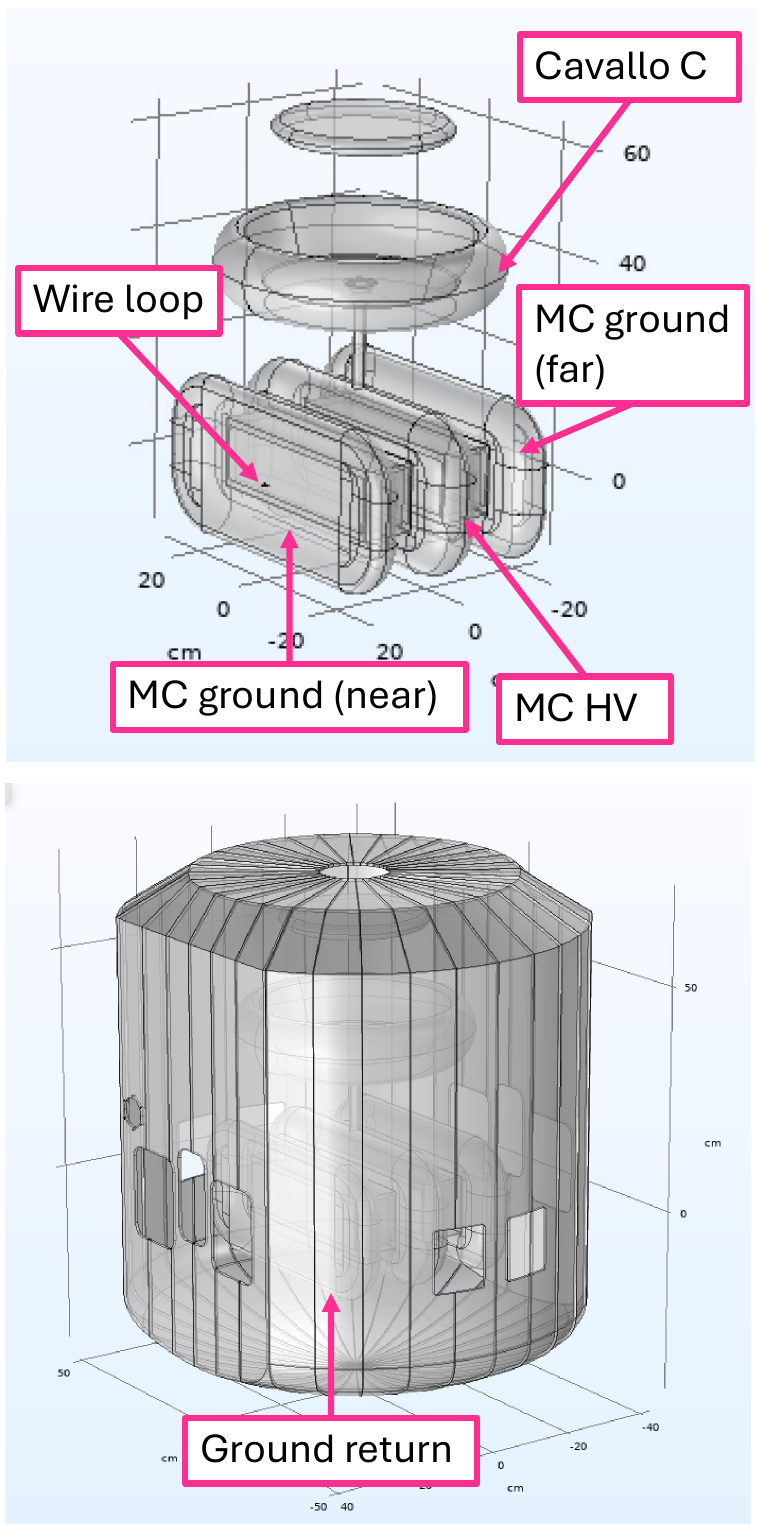}
    \caption{An example of COMSOL geometries used to calculate the magnetic Johnson noise from the electrodes at the SQUID loops using the method described in Ref.~\cite{Phan2024} based on the F-D theorem. Here ``MC'' stands for ``measurement cell''.}
    \label{fig:magnetic_Johnson_noise_COMSOL}
\end{figure}

We used this method to establish the resistivity requirements for the electrode materials. We built a model of the experiment in COMSOL, including the SQUID gradiometer loops, put an oscillating current in the SQUID gradiometer loops, and calculated the eddy current heating in each electrode for various resistivity values. The eddy current heating was related to the magnetic Johnson noise using Eq.~(\ref{eq:F-Dtheorem}). An example of the COMSOL models used for these evaluation is shown in Fig.~\ref{fig:magnetic_Johnson_noise_COMSOL}.

\begin{figure}
    \centering
    \includegraphics[width=1.0\linewidth]{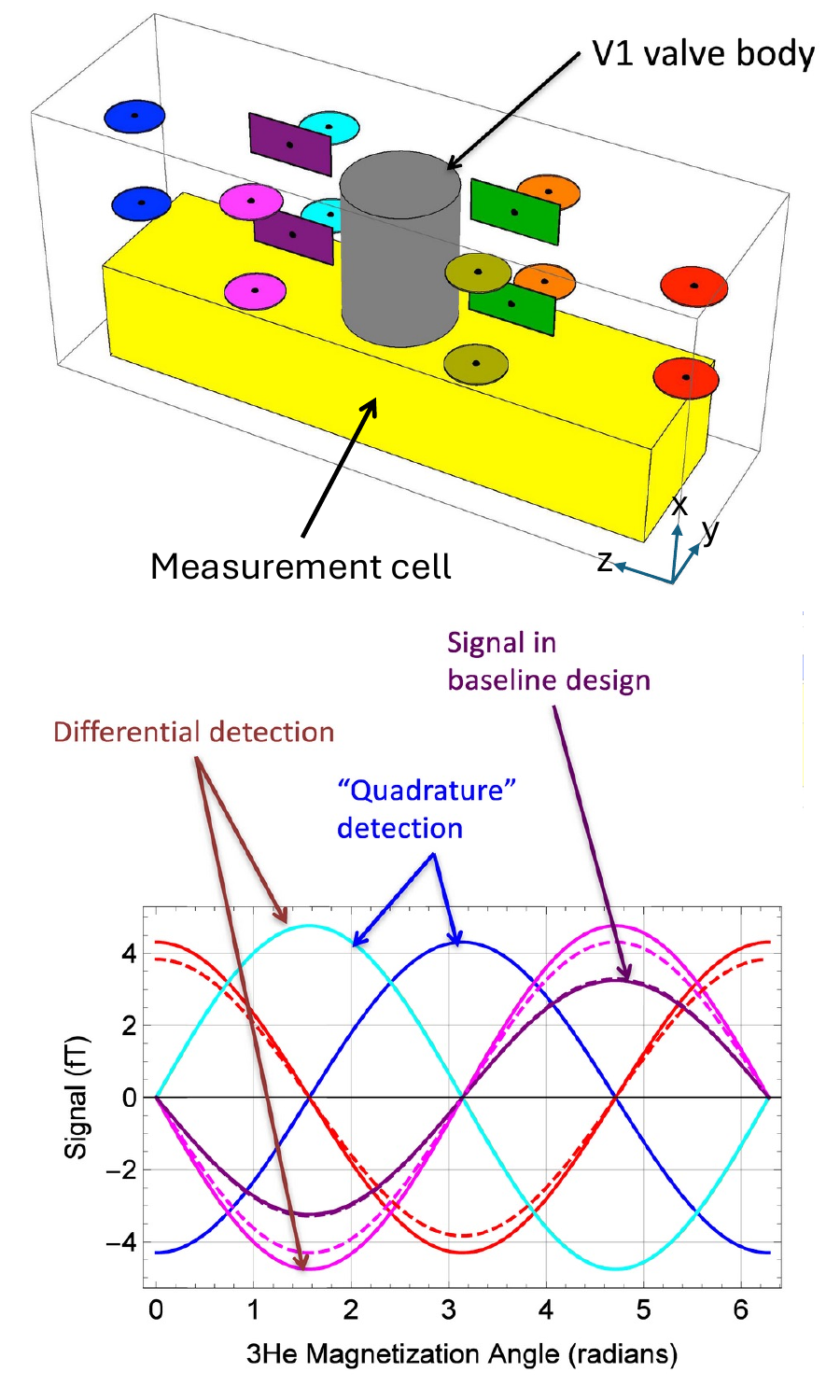}
    \caption{Top panel: SQUID pickup loop arrangement that use both axial (circles) and planar (rectangles) gradiometers. Bottom panel: expected magnetic field signal detected by each gradiometer. The same color is used for the expected signal and the corresponding gradiometer. }
    \label{fig:SQUIDloops}
\end{figure}
\begin{figure}
    \centering
    \includegraphics[width=1\linewidth]{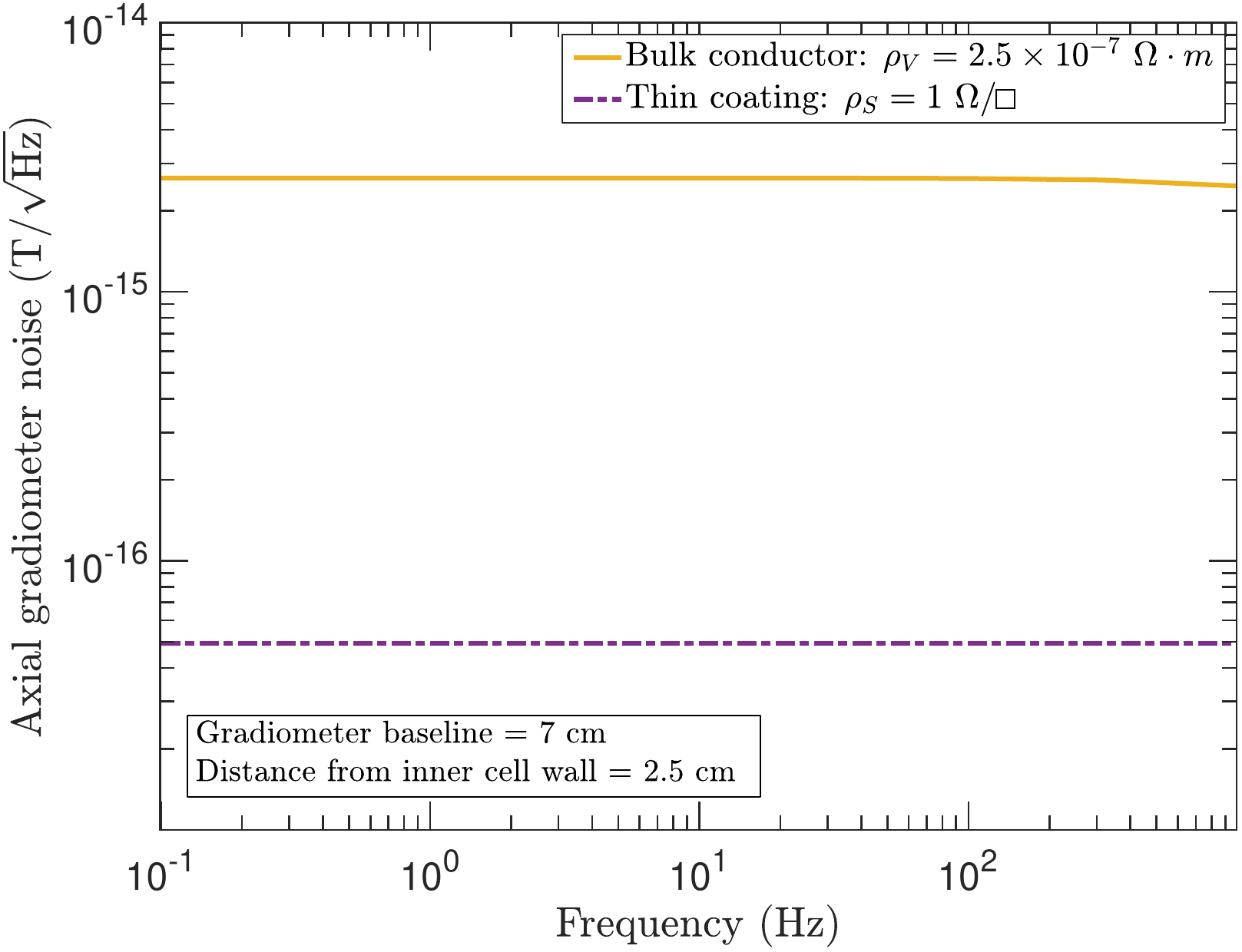}
    \caption{Calculated magnetic Johnson noise for axial SQUID gradiometers.}
    \label{fig:magnetic_Johnson_noise}
\end{figure}

In general, the magnetic Johnson noise is larger when the observation point is closer to the surface of a conducting material. For example, for an infinite plate, the magnitude of magnetic Johnson noise is inversely proportional to the distance between the observation point and the surface of the plane. For this reason, only the ground electrode adjacent to the SQUID loops makes non-negligible contributions to the magnetic Johnson noise. We evaluated the magnetic Johnson noise for several SQUID loop configurations (see Fig.~\ref{fig:SQUIDloops}) under two conditions: electrodes made of bulk conductors, and electrodes made of insulators coated with a conductive layer.  As an example, the calculated magnetic Johnson noise for the planar SQUID gradiometers is shown in Fig.~\ref{fig:magnetic_Johnson_noise}. From this analysis, we obtained the following requirements for the resistivity of the electrode materials: 
\begin{equation}
\rho_S \gtrsim 2\times 10^{-3} \;\;{\rm \Omega}/ \Box \;\;\;\; {\rm (for\;\;surface\;\;coating)}, 
\end{equation}
and
\begin{equation}
\label{eq:bulk_resitivity_Johnson_noise}
\rho_V \gtrsim 1.7\times 10^{-6} \;\;{\rm \Omega \cdot m} \;\;\;\; {\rm (for\;\;bulk\;\;materials)},    
\end{equation}
where $\rho_S$ and $\rho_V$ are the surface and volume resistivities, respectively. Note that the allowed value for $\rho_V$ depends on the details of the geometry, particularly the material thickness. The value shown in Eq.~(\ref{eq:bulk_resitivity_Johnson_noise}) assumes 5~mm thick material for the lobe part of the electrodes and 1~mm thick material for the flat part of the electrode. The requirements can be relaxed by thinning the flat part of the electrodes.

\subsubsection{Eddy current heating}
The resistivity requirements for eddy current heating from the dressing field were evaluated using a COMSOL model similar to that shown in Fig.~\ref{fig:magnetic_Johnson_noise_COMSOL}. In this case, the eddy current heating due to a uniform oscillating magnetic field that fills the entire volume (instead of an oscillating field due to an oscillating current flowing through a wire loop) was calculated. The ground return was segmented into 32 separated panels to reduce the eddy current heating, since the eddy current scales as the fourth power of the linear dimension of an object in the directions perpendicular to the field. The amplitude ($B_1=38$~$\mu$T) and frequency ($\omega_1 = 6000$~rad/s) of the dressing field were taken from Ref.~\cite{Swank2018}. We evaluated the eddy current heating under two conditions: electrodes made of bulk conductors, and electrodes made of insulators coated with a conductive layer. In both cases, the skin depth is much larger than the size of the electrode (or the thickness of the conductive layer). Therefore, the heating $P$ can be expressed in terms of $B_1$ and $\omega_1$ as follows:
\begin{equation}
\label{eq:eddy_current_heating_surface}
    P = F_{\rm geom}\frac{\omega_1^2 B_1^2}{\rho_s} = \frac{K}{\rho_S},
\end{equation}
and
\begin{equation}
\label{eq:eddy_current_heating_bulk}
    P = F_{\rm geom}' \frac{\omega_1^2 B_1^2}{\rho_v} = \frac{K'}{\rho_V},
\end{equation}
where $F_{\rm geom}$ and $F_{\rm geom}'$ are geometrical factor extracted from the results of COMSOL calculations performed for different values $B_1$ and $\omega_1$. As before, $\rho_V$ and $\rho_S$ are the volume and surface resistivities, respectively. The obtained values for $K$ and $K'$ are listed in Table~\ref{tab:eddy_current_heating}. Because of the large skin depth, the eddy current heating of different electrodes can be calculated independently. From these results and the requirement that the eddy current heating needs to be less than 6~mW, we obtained the following requirements for the resistivity of the electrode materials: 
\begin{equation}
\rho_S \gtrsim 0.013\;\;{\rm \Omega}/ \Box \;\;\;\; {\rm (for\;\;surface\;\;coating)}, 
\end{equation}
and
\begin{equation}
\rho_V \gtrsim 1.42\times 10^{-4} \;\;{\rm \Omega \cdot m} \;\;\;\; {\rm (for\;\;bulk\;\;materials)}.   
\end{equation}

\begin{table}[t]
    \centering
    \begin{tabular}{ccc} \hline \hline
Electrode         &  $K$ (W $\cdot \Omega$) & $K'$ (W$\cdot \Omega \cdot$m)\\  \hline
Ground return  & $2.47\times 10^{-5}$ & $1.69\times 10^{-7}$\\ 
Cavallo C & $2.40\times 10^{-5}$ & $3.66\times 10^{-7}$ \\ 
Cavallo B & $8.77\times 10^{-7}$ & $4.53\times 10^{-9}$\\ 
MC HV & $1.22\times 10^{-5}$ & $1.65\times 10^{-7}$\\ 
MC ground & $1.40\times 10^{-5}$ & $1.44\times 10^{-7}$\\ \hline
Total & $7.57\times 10^{-5}$ & $8.49\times 10^{-7}$ \\ \hline\hline
\end{tabular}
    \caption{Numerical values for the coefficients $K$ and $K'$ in Eqs.~(\ref{eq:eddy_current_heating_surface}) and (\ref{eq:eddy_current_heating_bulk}).}
    \label{tab:eddy_current_heating}
\end{table}

\section{Relevant physical phenomena}
\label{sec:physical_phenomena}

\subsection{Electrical breakdown in liquid helium}
\label{subsec:electrical_breakdown}
At the time when our R\&D effort started, data existed on electric breakdown in LHe for various electrode geometries---such as sphere to sphere, sphere to plane, and plane to plane---in the temperature range of $1.2-4.2$~K, and mostly at the saturated vapor pressure (SVP) (see, e.g., Ref.~\cite{Gerhold1998} for a review of earlier work). While the existing data showed little consistency in general, several parameters were identified as potentially affecting breakdown, including (1) electrode material and surface finish, (2) gap size and/or electrode size, and (3) temperature and/or pressure of the liquid. In earlier studies, while a general dependence of the breakdown on the size of the system was recognized, it was not clearly determined whether the breakdown depended on the area of the electrode, the gap size, or the size of the stressed volume. 
This ambiguity arose because those studies failed to account for the fact that varying the gap size for the same set of electrodes also changes the stressed area on the electrodes.
Similarly, those studies were not able to differentiate the effects of temperature and pressure because the temperature was varied by evaporatively cooling the LHe in which the electrodes were immersed. Note that the SVP of LHe changes from 760~torr to 0.3~torr as the temperature changes from 4.2~K to 1.2~K. 

Based on a detailed theoretical study of electron multiplication induced by an electric field in LHe starting from the kinetic Boltzmann equation~\cite{Belevtsev1993}, the intrinsic breakdown field of LHe is expected to be at least in the low~MV/cm region. On the other hand, breakdown fields around or below a few 100~kV/cm have been commonly reported in the literature for LHe. These observations are therefore presumed to be the result of the following process: (1)~Field emission of electrons occurs on a rough surface because of extremely high local fields at sharp asperities. (2)~This causes local heating, which results in vapor bubble formation, leading to electron multiplication and vapor growth proceeding together. (3)~Breakdown is then the result of a vapor column extending to the anode and the gas within the column undergoing electrical breakdown. Note that in other dielectric liquids, formation and growth of vapor bubbles associated with electrical breakdown have been observed experimentally~\cite{Kelley1981,Kattan1989,Forster1990}.

In Ref.~\cite{Phan2021}, we reported a comprehensive study of electrical breakdown in LHe performed using a set of small near-uniform field stainless steel electrodes (electrode size $\sim$1~cm with a stressed area of $\sim$0.7~cm$^2$). We measured the breakdown voltage repeatedly to obtain statistical distributions of the breakdown voltage for different electrode finishes (electropolished and mechanically polished) at temperatures between 1.7 and 4.0~K and pressures between the SVP and 626 Torr. Here we summarize the major findings relevant to the discussions in the subsequent sections, as these points provide essential context for the analysis presented later in this paper. 
\begin{figure*}
    \centering
    \includegraphics[width=1.0\linewidth]{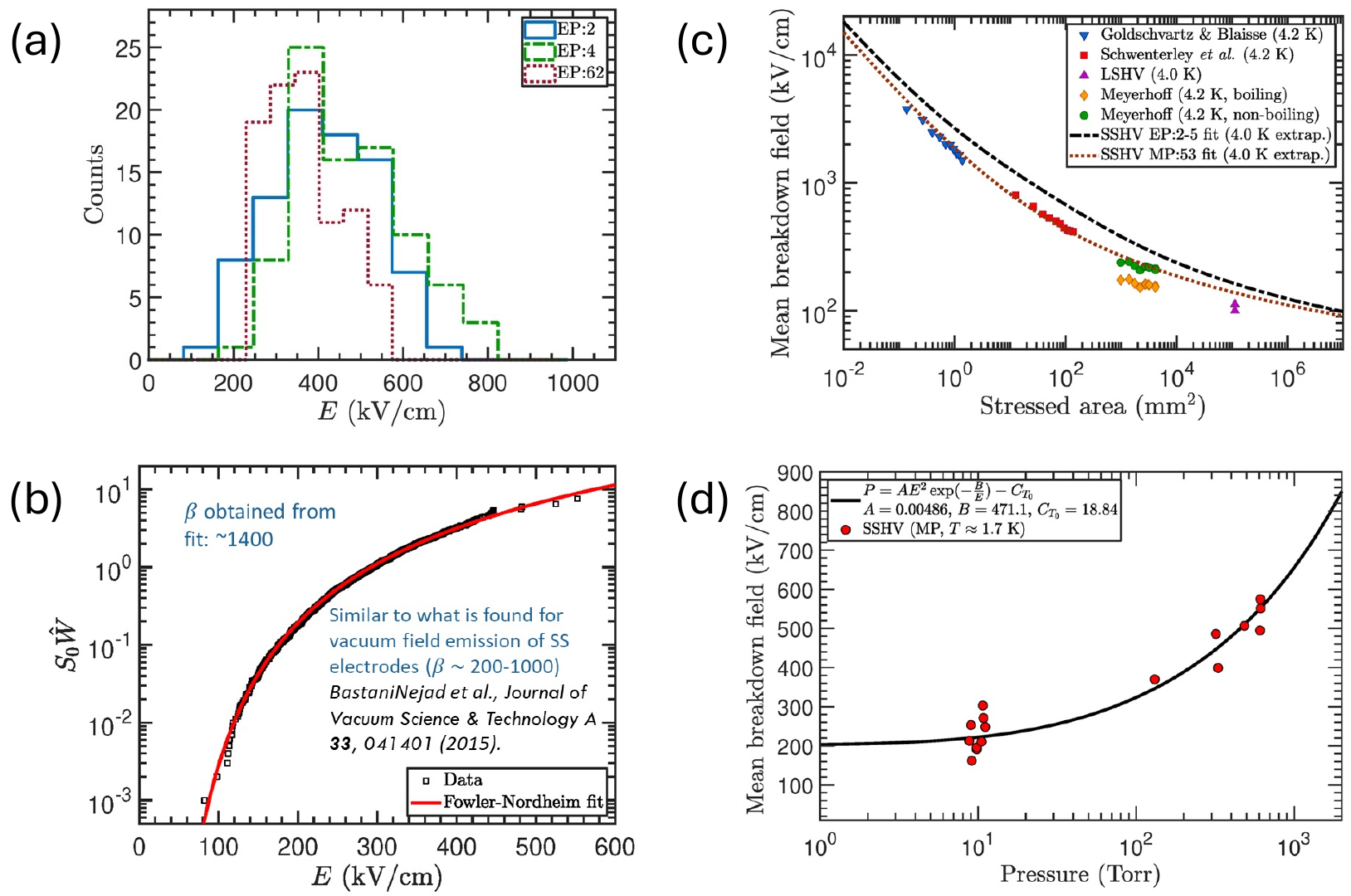}
    \caption{(a) A plot showing typical breakdown voltage distributions measured in Ref.~\cite{Phan2021}. (b)~Hazard function obtained from the measured breakdown voltage distribution. The hazard function is well described by the Fowler-Nordheim equation for field emission. (c)~The hazard function determined from the measured breakdown field distribution can be used to predict the breakdown field distribution for electrodes of different sizes. The prediction made this way is compared to data in the literature, showing an excellent agreement. (d)~The breakdown field depends on the pressure of the LHe.}
    \label{fig:sshv}
\end{figure*}

The HV ramped at a constant rate of 50~V/s, 100~V/s, or 200~V/s until a breakdown was detected, which then tripped the power supply. The voltage at which the breakdown was detected was recorded and the same process was repeated. An example of the breakdown voltage distributions thus obtained is shown in Fig.~\ref{fig:sshv}~(a). No dependence on the ramp rate was observed. Based on the most general formulation for breakdown statistics~\cite{Choulkov2005}, the obtained breakdown distribution, once cast in the form of the cumulative breakdown distribution $P_b(E)$ (the probability that breakdown occurs at or below electric field $E$), can be expressed in the following form:
\begin{equation}
    \label{eq:cumulative_breakdown_distribution}
    P_b(E) = 1- e^{-S_0 W(E)},
\end{equation}
where $S_0$ is the stressed area of the electrodes and $W$ is the hazard function. The value of $S_0 W(E)$ can readily be extracted from the data. Figure~\ref{fig:sshv}~(b) shows an example of the $S_0 W(E)$ extracted in this way for mechanically polished stainless steel electrodes. We found that the curve of $S_0 W(E)$ is well fitted with the Fowler-Nordheim equation for the field emission current $I$
\begin{equation}
    I = A_e \frac{1.54}{\phi}10^{4.52\phi^{-1/2}}(\beta E)^2 \exp \left (\frac{-6.53\times 10^4 \phi^{3/2}}{\beta E} \right )\;{\rm A},
\end{equation}
where $A_e$ is the effective emission area in cm$^2$, $\beta$ is the local field enhancement factor, $\phi$ is the electrode work function in eV, and $E$ is the applied electric field in kV/cm. The empirically obtained hazard function $S_0 W(E)$ is well fitted with the Fowler-Nordheim equation with $\beta \sim 1400$ [see Fig.~\ref{fig:sshv}~(b)], which is similar to what is found in vacuum field emission from stainless steel electrodes of $\beta \sim 200-1000$. This connection between the hazard function for electrical breakdown and the Fowler-Nordheim equation for field emission supports the above mentioned presumed process that leads to electrical breakdown in LHe. Therefore, in Eq.~(\ref{eq:cumulative_breakdown_distribution}), the electrode area $S_0$ is taken to be that of the cathode.

We compared the hazard function obtained for different electrode finishes, specifically between mechanically polished and electropolished stainless steel surfaces. We found that the hazard function for the mechanically polished stainless steel agrees well with the hazard function for the electropolished stainless steel multiplied by a factor of 7. This suggests that the surfaces of our mechanically polished stainless steel sample have a 7 times higher density of breakdown initiation sites compared to those of our electropolished stainless steel samples. Electropolished stainless steel has been empirically demonstrated to be an excellent electrode material. In contrast, mechanical polishing, while reducing surface roughness at the macroscopic scale, introduces numerous microscopic asperities that can serve as initiation sites for electrical breakdown. While stainless steel was clearly unsuitable as the electrode material for the Cryogenic nEDM experiment due to its magnetic properties and high electrical conductivity, it was expected that electropolished and mechanically polished stainless steel surfaces—representing two extremes of surface treatment—would provide an indication of the possible range of breakdown behavior of other materials.

From Eq.~(\ref{eq:cumulative_breakdown_distribution}), it is straightforward to determine how the breakdown probability scales with electrode area for a near-uniform electrode whose area is $n$ times larger than that used in this measurement, namely,
\begin{equation}
    P_b(E,n) = 1 - e^{nS_0 W(E)}.
\label{eq:scaledcdf}
\end{equation}
Figure~\ref{fig:sshv}~(c) shows a comparison between the predicted area scaling behavior based on our data from Ref.~\cite{Phan2021} and Eq.~(\ref{eq:scaledcdf}) and data that had been taken earlier by other researchers~\cite{Goldschvartz1966,Meyerhoff1995,Schwenterly1974}. 
The agreement between these data and our prediction based on our data is remarkable, especially considering that the measurements were carried out using different apparatus and by different researchers. The fact that the electrode gap sizes were different for the various measurements further reconfirms that the electric field on the electrode surface, rather than the voltage difference between electrodes, is the parameter relevant to describing electrical breakdown. This, in turn, is consistent with the above-mentioned picture in which breakdown is initiated on the electrode surface.
Furthermore, it is seen that the area scaling---namely, the observation made not only with LHe but also with other dielectric fluids that the breakdown field is lower for a larger system---follows naturally from our treatment. In Ref.~\cite{Phan2021} we also presented an analysis of existing data on electrical breakdown in liquid argon (LAr). A hazard function was obtained from data taken with a set of electrodes (and one stressed area value). The breakdown field calculated for other stressed electrode area values based on the obtained hazard function reproduces the data very well, indicating that this method applies also to LAr and possibly other liquids. 

The area scaling of Eq.~(\ref{eq:scaledcdf}) can be further extended to the general case in which the electrode has an arbitrary shape and an arbitrary field distribution. That is,
\begin{equation}
    P_b = 1 - \prod_{\{ E_i \}} e^{-S(E_i)W(E_i)}.
\label{eq:scaledcdf_gen}
\end{equation}
Here $S(E_i)$ is the area of the electrode that has a field value $E_i$ and a limit in which $E_i$ takes continuous value is assumed. The fact that the breakdown probability for an arbitrary shaped electrode is expressed in terms of the hazard function $W(E)$---which can be determined through a measurement that uses a set of small near-uniform electrodes---provides an efficient path for electrode development. That is, once a candidate electrode material is identified, one can make a measurement similar to that shown in Fig.~\ref{fig:sshv}~(a) using a small sized cryostat. An electrode system of required features can be simulated using a finite element analysis tool such as COMSOL~\cite{COMSOL}. The expected performance, in terms of breakdown probability, can be computed using Eq.~(\ref{eq:scaledcdf_gen}) along with the obtained field distribution. 

We observed that the breakdown field depends on the pressure of the LHe, as shown in Fig.~\ref{fig:sshv}~(d). This also is consistent with the above mentioned mechanism for electrical breakdown in LHe, as a higher pressure can suppress the growth and propagation of vapor bubbles in liquid. 

As is evident from the derivation, the probability of breakdown given in Eqs.~(\ref{eq:cumulative_breakdown_distribution}), (\ref{eq:scaledcdf}), and (\ref{eq:scaledcdf_gen}) represents the probability of breakdown per voltage ramp. In this regard, we also reported in Ref.~\cite{Phan2021} a study of the breakdown probability for successive voltage ramps. We found that if a breakdown does not occur below a voltage $V_0$ during the first ramp and the voltage is subsequently ramped down before breakdown occurs, then no breakdown occurs below $V_0$ during the second ramp. Implications of this observation is discussed in Sec.~\ref{sec:expected_performance}.

\subsection{Effects of cosmic rays and other ionizing radiation}
\label{subsec:cosmic_rays}
Whether cosmic rays or other ionizing radiation can initiate breakdown is an important question for the design of the HV system in the Cryogenic nEDM experiment. Here we provide direct experimental evidence, from our own data and that of others, supporting the inference that such radiation does not induce breakdown, at least in the electric field range of interest.

We discussed earlier that an electric field at least in the MV/cm
region is needed to create a breakdown from a free electron in LHe by
accelerating it and producing an avalanche~\cite{Belevtsev1993} and that the electrical breakdown in LHe observed to occur at much lower fields is thought to result from electrical breakdown through a column of gas extending from the cathode to the anode. Therefore, in order for a cosmic ray particle or other ionizing radiation to produce a breakdown, a sufficient energy density has to be deposited so that a column of gas is created in the liquid that extends from one electrode to the other.

As discussed in Ref.~\cite{Ito2012}, an $\alpha$ particle with a kinetic
energy of 5.5~MeV (which has a much larger linear energy transfer than
cosmic ray particles) has a range of 0.3~mm in LHe. The produced
secondary electrons stop within 60~nm from the primary
track. Therefore the 5.5~MeV energy is deposited in a column with a
length of 0.3~mm and a radius of 60~nm.  This corresponds to a
deposited energy density of 7~J/mol, not sufficient to create a gas
filled column vaporizing LHe. At 0.4~K and at the saturated vapor
pressure ($10^{-6}$~Torr), the latent heat of
vaporization is 68~J/mol~\cite{Donnelly1998}. At 4.2~K and at the saturated
vapor pressure (760~Torr), the latent heat of vaporization is
83~J/mol~\cite{Donnelly1998}. If the liquid is at 0.4~K and is pressurized to
760~Torr (the SVP for 4.2~K), it requires additional 39~J/mol to increase the temperature of the liquid from 0.4~K to 4.2~K~\cite{Donnelly1998}. An energy deposition of 7~J/mol can raise the temperature of the LHe in the column only to
$\sim$2~K. Note that the reaction products from neutron capture on $^3$He deposit a smaller energy density than these alpha particles~\cite{Ito2012}. Minimum ionizing particles, such as cosmic ray muons or fast electrons, produce tracks with a still smaller energy density. 

In experiments to study the effect of an electric field on LHe
scintillation due to different types of ionizing
radiation~\cite{Ito2012,Phan2020} (performed as part of the R\&D for
the Cryogenic nEDM experiment), we electroplated radioactive sources on an electrode. In
Ref.~\cite{Ito2012}, approximately 300~Bq of $^{241}$Am (which emits
5.5~MeV $\alpha$'s) was electroplated on the ground electrode. The
activity had a diameter of 6~mm. An electric field of up to 45~kV/cm
was applied to the gap between the HV and ground electrode. Electrical
breakdowns were observed when increasing the voltage to reach a new
voltage value for the first time. However, once the target voltage
value was reached, we did not observe electrical breakdown for the
duration of the measurement (at each voltage setting, data were
acquired for 5 min). In this experiment, the HV ground was biased with
a negative voltage therefore the ground electrode was the anode. 
The non-observation of electrical breakdown in the presence of a radioactive source in this experiment is consistent with the analysis presented above.

In Ref.~\cite{Phan2020}, we co-electroplated $^{113}$Sn (which emits
364-keV conversion electrons) and $^{241}$Am in a 6.35-mm spot on the
HV electrode. The $^{113}$Sn and $^{241}$ Am sources had activities
that corresponded to emission rates of 850~s$^{-1}$ and 195~s$^{-1}$
for 364- to 391-keV electrons and 5.388- to 5.544-MeV $\alpha$
particles, respectively, into the liquid at the time of the
experiment. An electric field of up to 40~kV/cm was applied to the gap
between the ground and HV electrodes. Both positive and negative
polarities were used. We did not observe any electrical breakdown in
this experiment either. 

Reference~\cite{Long2006} reports a test in which the breakdown voltage
was measured with and without a radioactive source placed near a system in which a set of 50~cm diameter electrodes separated by $\sim$3~mm was immersed in LHe and was subject to a potential difference of $\sim$30~kV. A $1.1\times 10^{11}$~Bq PuBe source was used for this test. 
The breakdown voltage was measured both with and without the PuBe source, and no reduction attributable to the source was observed. This result is again consistent with our earlier analysis, noting that the $\alpha$ particles generated by 1-MeV neutrons recoiling off $^4$He nuclei have much smaller energies than 5.5-MeV $\alpha$ particles. The flux of neutrons in LHe due to the $1.1\times 10^{11}$~Bq PuBe source was comparable to that expected for the neutron flux in the measurement cells in the Cryogenic nEDM experiment, although the energies of the neutrons were much different (1~MeV vs 1~meV).

\section{Design and demonstrated component level performance}
\label{sec:design}
\subsection{Overview}
The task of developing the HV and electrode system for the Cryogenic EDM was divided into several mutually related subtasks. They are: (1)~the development of the measurement cell electrodes, (2)~the development of a method for delivering a sufficiently high electric potential to the measurement cell electrodes, and (3)~identification of appropriate electrode materials. We describe the design and demonstrated performance of each below.

\subsection{Measurement cell electrodes}
\begin{figure}
    \centering
    \includegraphics[width=0.9\linewidth]{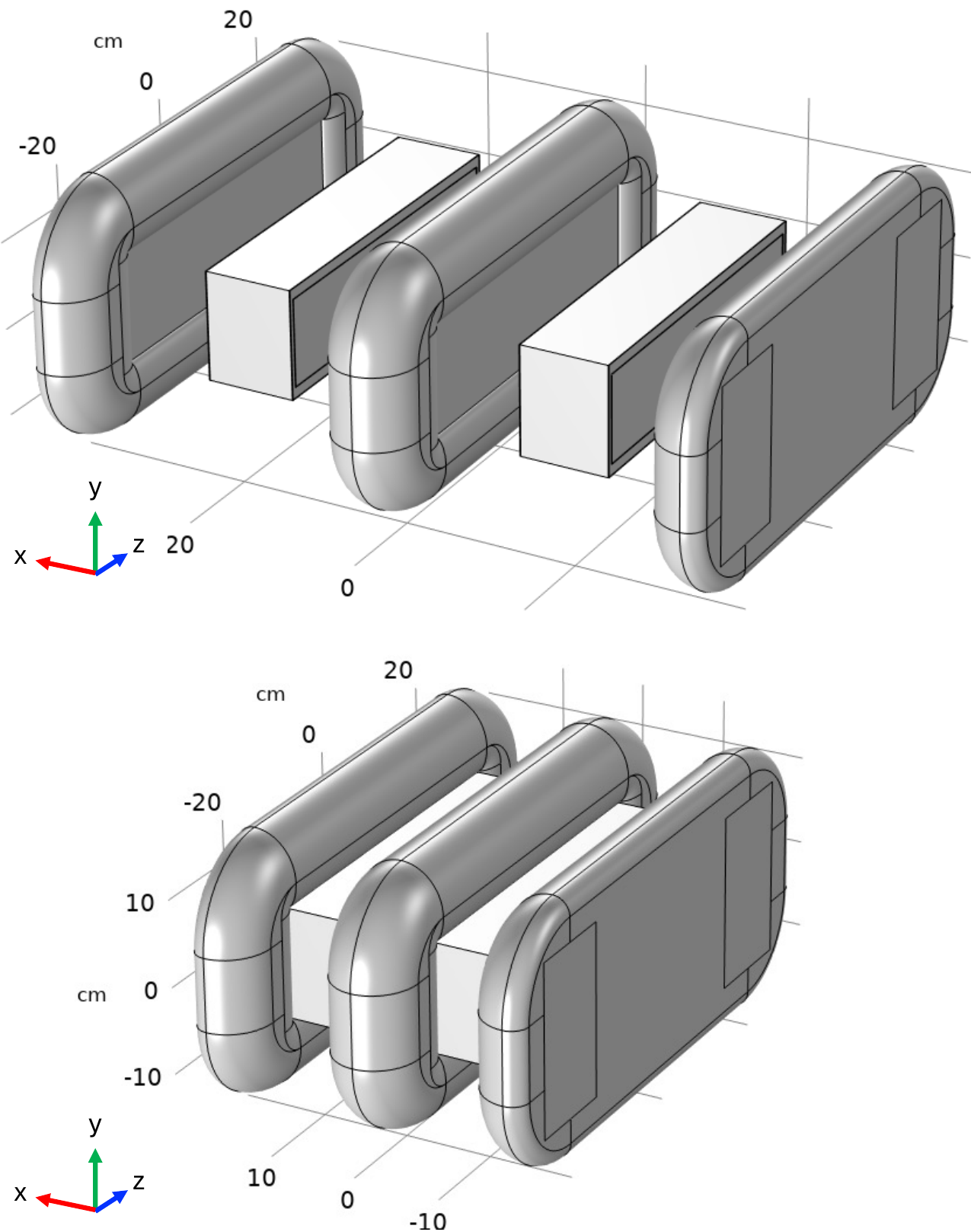}
    \caption{Design of the Measurement Cell electrodes}. The upper panel shows the individual components. The lower panel shows the electrode-cell assembly. 
    \label{fig:MC_electrode_design}
\end{figure}

The shape of the measurement cell electrodes was designed using COMSOL~\cite{COMSOL} with the following considerations in mind:
\begin{itemize}
    \item The edges of the measurement cells need to be embedded in a recess on the electrode to avoid  the enhancement of the field at the cathode-insulator junction due to the higher dielectric constant of the dielectric measurement cell materials (see, e.g., Ref.~\cite{Ito2016} and references therein). 
    \item This in turn means that both the HV and ground electrodes will have a lobe surrounding a flat region that faces one of the faces of the measurement cell. And the surface of the lobe will have a much higher electric field than that inside the measurement cell. We required that the highest field on the surface of the lobe when the field inside the measurement cell is 75~kV/cm be less than 120 kV/cm based on findings from our earlier work~\cite{Ito2016}. The ideal procedure to optimize the shape of the electrode would be to use Eq.~(\ref{eq:scaledcdf_gen}) with the hazard function $W(E)$ of the chosen electrode material to calculate the breakdown probability for each trial geometry. However, to make the process more efficient, we chose to design the electrode initially with this simple criterion and subsequently verify the resulting design using Eq.~(\ref{eq:scaledcdf_gen}).
\end{itemize}
In COMSOL, the electrodes were constructed from various COMSOL primitives and surfaces generated by revolving a two-dimensional curve. Several parameters, including the radii of curvature of different parts of the lobe, were varied until a satisfactory solution was obtained. The resulting design is shown in Fig.~\ref{fig:MC_electrode_design}.

\begin{figure}
    \centering
    \includegraphics[width=0.9\linewidth]{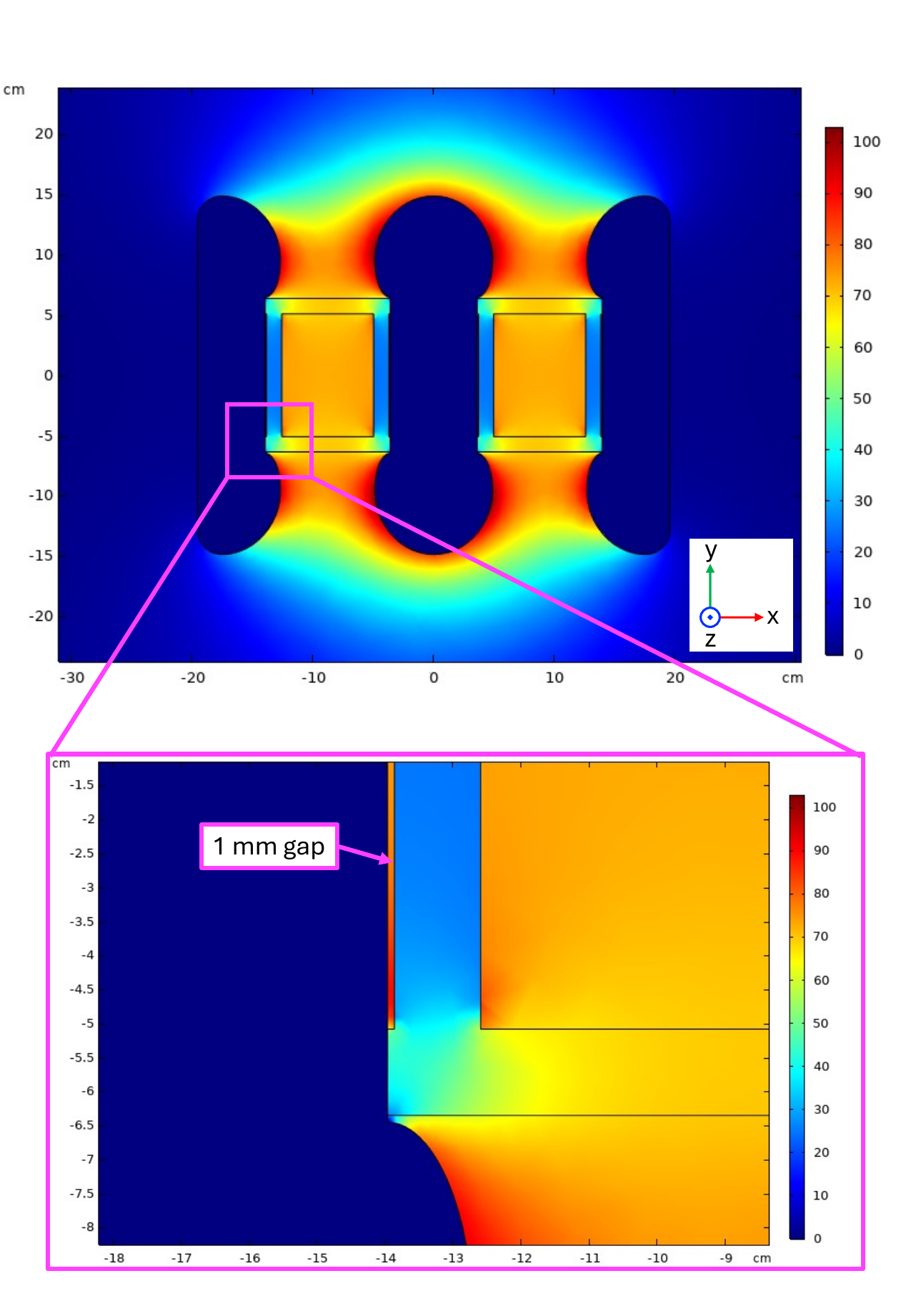}
    \caption{Electric field distribution calculated using COMSOL for the electrode geometry shown in Fig.~\ref{fig:MC_electrode_design}.}
    \label{fig:MC_electric_field}
\end{figure}

Figure~\ref{fig:MC_electric_field} shows the calculated electric field distribution when 635~kV was applied to the HV electrode. The lower panel of the figure shows the electric field distribution near an edge of the measurement cell. 
The edges of the measurement cells are in a region with low electric fields because of the recess on the electrode.
The lower panel of the figure also shows a 1-mm gap between the electrode and the measurement cell. This gap is required to provide a sufficient thermal contact between the liquid helium and the measurement cell surfaces, which is required for moving $^3$He atoms using the heat flush method~\cite{Ahmed2019}.

\subsection{Cavallo's multiplier for HV generation}
\begin{figure*}
    \centering
    \includegraphics[width=1\linewidth]{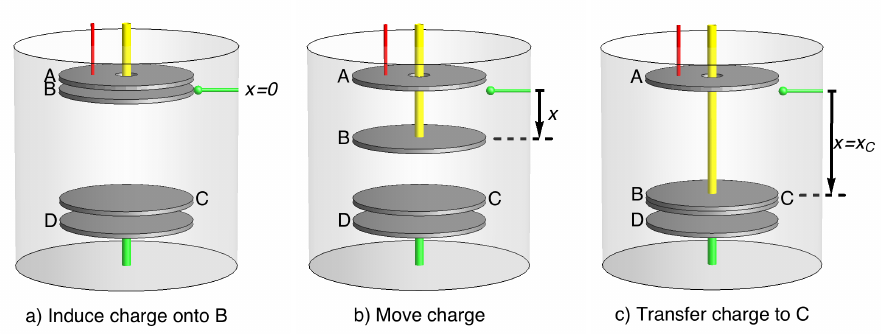}
    \caption{Diagrams showing the principle of Cavallo's multiplier. a)~Electrode A is connected to a power supply that provides modest HV (e.g., 50~kV). Electrode B is initially grounded. Charge is induced electrostatically due to the capacitance between Electrodes A and B. Electrode C is the electrode to be charged up to a high voltage (e.g., 635~kV). Electrode D is a ground electrode. b)~Electrode B is pushed downward. As soon as it starts to move, it is disconnected from the ground. Now, Electrode B carries the charge that was on it in a) .c)~Electrode B comes into contact with Electrode C. The charge on Electrode B moves to Electrode C because of the capacitance between C and D. A finite fraction of charge remains on B due to the capacitance between the back of B and the surrounding ground. By repeating a)--c), it is possible to accumulate charge on C, raising C to a high voltage. The maximum voltage achievable is given by 
    $V_C^{\rm max} \approx V_A (C_{AB}^0 - C_{AB}^1)/(C_{BC}^0+C_{AB}^1+C_{BG}^1)$, where the subscripts on $C$ indicate the electrode pair defining the capacitance, and the superscript denotes the location of Electrode B with ``0'' meaning B is located near A and is grounded and ``1'' indicating that it is in contact with C~\cite{Clayton2018}.}
    \label{fig:Cavallo_principle}
\end{figure*}

Cavallo's multiplier~\cite{Cavallo1795} is an electrostatic induction machine that generates HV by repeatedly transferring electrostatically induced charge to the HV electrode. It is ideally suited for producing the required 635 kV in the cryogenic nEDM experiment because it can be implemented within the same LHe volume as the measurement cells and their electrodes. This design eliminates the need for a direct 635 kV HV feed, which would be extremely difficult to make compatible with several of the experiment’s technical requirements, including heat load, magnetism, and SQUID operation. With Cavallo's multiplier, it is sufficient to feed a modest high voltage of about 50~kV, which can easily be implemented in a manner compatible with other experimental requirements. In addition, Cavallo's multiplier can be implemented much more compactly than the variable capacitor method~\cite{Long2006}, which was also considered as an option for in situ voltage generation. Figure~\ref{fig:Cavallo_principle} shows the principle of voltage amplification of Cavallo's multiplier. 

We presented our initial analysis of the application of Cavallo's multiplier to  the Cryogenic nEDM experiment and discussed its unique suitability in Ref.~\cite{Clayton2018}. Building on that work, we designed a full-scale Cavallo electrode system that has a geometric gain of 18 (the gain solely limited by the electrostatics as described in the caption of Fig.~\ref{fig:Cavallo_principle} and not by electric breakdown) and a sufficiently small breakdown probability when Eq.~(\ref{eq:scaledcdf_gen}) was evaluated using the hazard function for electropolished stainless steel~\cite{Blatnik2025}. 

\begin{figure}
    \centering
    \includegraphics[width=1.0\linewidth]{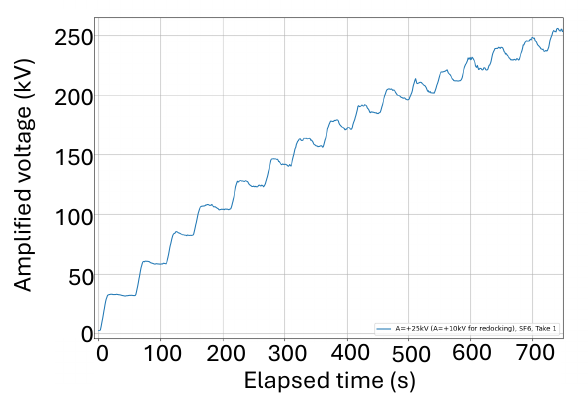}
    \caption{Cavallo charging curve obtained by running the Cavallo system in SF$_6$, demonstrating that 250~kV was achieved.}
    \label{fig:Cavallo_250kV}
\end{figure}

Following the design stage, we constructed a full-scale system with stainless steel electrodes, and we tested it performance at room temperature using SF$_6$ gas as the insulating medium. We successfully demonstrated the generation of 250~kV from a 25-kV input voltage, with the limitation coming from the insulating performance of SF$_6$ and the surface finish of the HV electrodes~\cite{Clayton2025}. A charging curve demonstrating 250~kV is shown in Fig.~\ref{fig:Cavallo_250kV}.

Subsequently we proceeded to test the Cavallo apparatus cryogenically, using liquid nitrogen (LN$_2$) as the insulating medium. Using the same apparatus employed to obtain the results presented in Ref.~\cite{Phan2021}, we compared the electrical breakdown properties of liquid helium (LHe) and LN$_2$, and confirmed that LN$_2$ exhibits breakdown characteristics similar to those of pressurized LHe~\cite{Phan2025}. In addition, thermal contraction—the primary mechanical concern when designing and constructing cryogenic systems with moving parts—occurs predominantly between room temperature and LN$_2$ temperature, with very little additional contraction between LN$_2$ and LHe temperatures. Therefore, both mechanical and high-voltage amplification performance tests can be carried out using LN$_2$, which is technically simpler and substantially less costly than using LHe. We confirmed that the actuator performs without problem at the LN$_2$ temperature. We also confirmed voltage amplification. 

\subsection{Electrode materials}
The resistivity requirements presented in Sec.~\ref{sec:resistivity_requirements} refer to the resistivity at 0.4~K. Therefore, these requirements immediately exclude pure metals: their resistivities are already too low at room temperature (typically in the $10^{-8}$~$\Omega\cdot$m range) and they decrease by orders of magnitude by going to 0.4~K. The resistivity of semiconductors tends to increase as the temperature is lowered. While most semiconductors become completely insulating at 0.4~K, some remain conductive. The resistivity of alloys tends to show little temperature dependence. Although the resistivity of most alloys is too low, some have values within the acceptable range. An extensive materials search and R\&D effort identified the following three candidate materials: (1) PMMA-coated copper germanium (Cu-Ge), (2) silicon bronze, and (3) silicon carbide. The first two are alloys and the third one is a semiconductor. 

\begin{figure}
    \centering
    \includegraphics[width=0.9\linewidth]{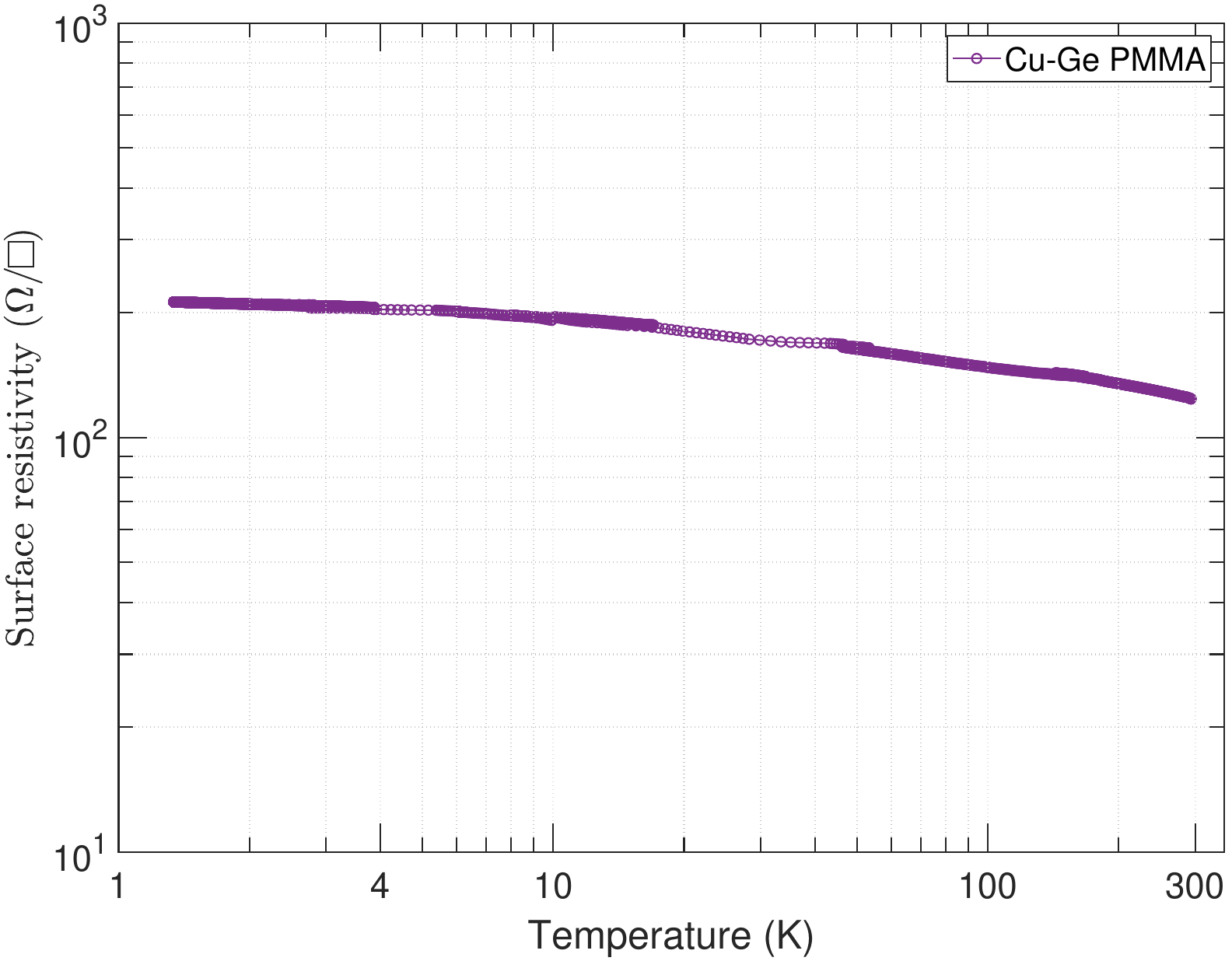}
    \caption{Surface resistivity of Cu-Ge coated PMMA sample as a function of the temperature.}
    \label{fig:CuGe_RvsT}
\end{figure}

Amorphous copper germanium (a-Ge$_{1-x}$Cu$_x$) is an alloy whose resistivity can be adjusted by changing the copper concentration $x$ and when made into a thin film of $150-200$~nm, it exhibits a resistivity in the range that meets our requirements~\cite{Aboelfotoh1993,Wong1993}.  Low-magnetic-Johnson-noise thin-metal-film-based electrodes have been used for electron EDM experiments previously~\cite{Rabey2016}. Both copper and germanium, when used as a thin film, have acceptable neutron absorption properties. We successfully produced Cu-Ge coated PMMA samples and measured the surface resistivity as a function of the temperature. We confirmed that the resistivity is in the acceptable range as shown in Fig.~\ref{fig:CuGe_RvsT}.

Silicon bronze is an alloy composed of mostly copper and silicon. Typically it is over 92\% copper. The rest is silicon (typically $3-4$\%) and other elements such as zinc, lead, iron, manganese and nickel. Among the different types available, we evaluated Herculoy (CDA876) silicon bronze. The resistivity of silicon bronze and its temperature dependence are well documented in the literature. The resistivity at room temperature is $\rho_v (273\;{\rm K}) = 2.46\times 10^{-7}$~$\Omega\cdot$m. It changes only by 10\% between room temperature and 2.2~K. While this resistivity is smaller than what is shown in Eq.~(\ref{eq:bulk_resitivity_Johnson_noise}), the magnetic Johnson noise can be made acceptable by fabricating the flat part of the electrodes using a foil. 
Therefore this resistivity is compatible with the free precession measurement mode of the experiment. We measured its magnetism using fluxgate magnetometers in a magnetically shielded room. We did not detect any sign of magnetism at the level of $\sim$0.2~nT at 1~cm from the sample. The effective of the neutron absorption was evaluated though an analysis based on simulations. The predicted background rate was determined to be acceptable. 

\begin{figure}
    \centering
    \includegraphics[width=1.0\linewidth]{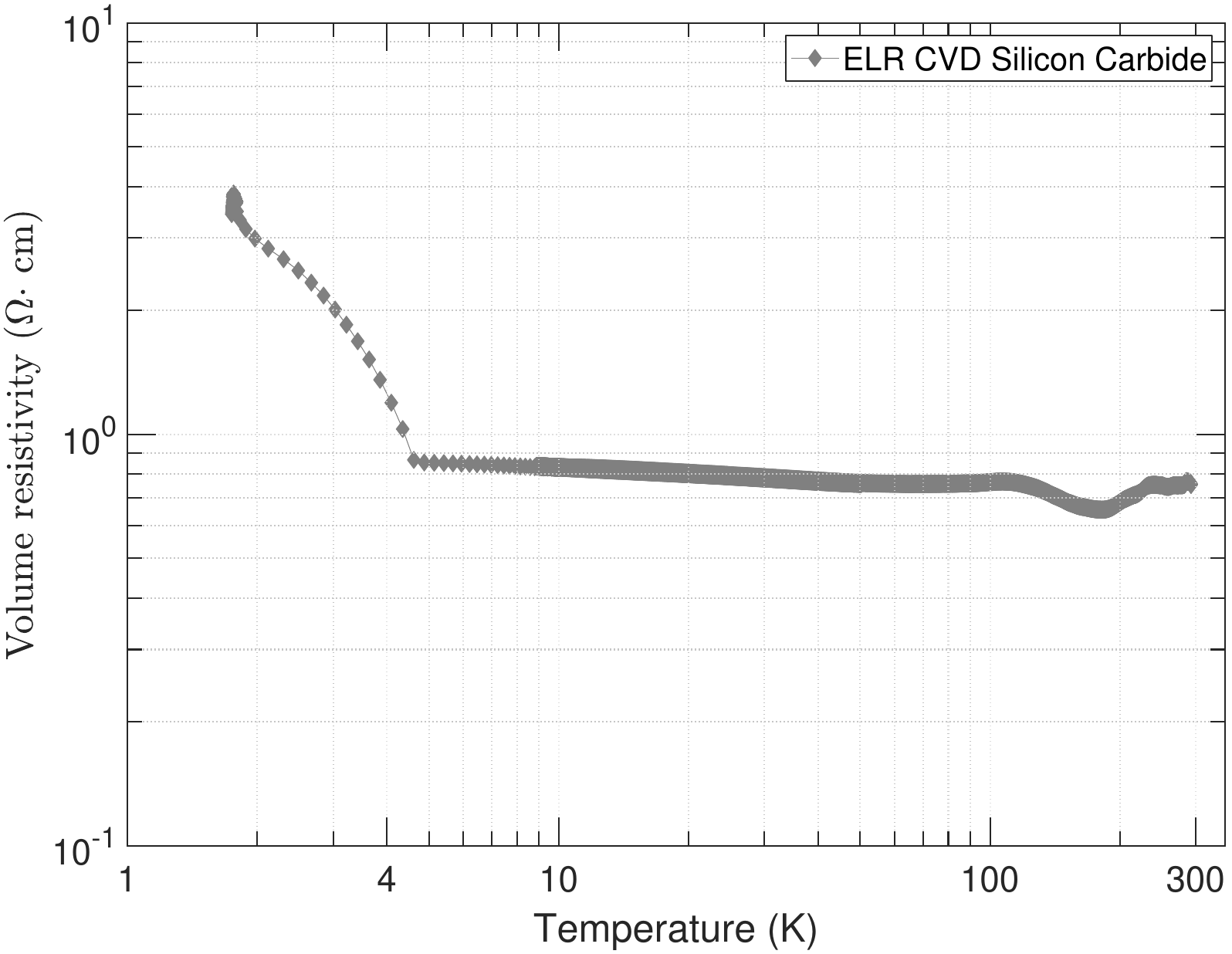}
    \caption{Resistivity of SiC measured as a function of the temperature}
    \label{fig:SiC_RvsT}
\end{figure}

Silicon carbide (SiC) is a compound containing silicon and carbon. As a semiconductor, its resistivity increases as the temperature is reduced. We measured the resistivity as a function of temperature for different types of SiC and found that one of them [extremely low resistivity (ELR) chemical vapor deposition (CVD)] meets the requirements for both the free precession method and the critical dressing methods. Figure~\ref{fig:SiC_RvsT} shows the measured resistivity of SiC as a function of the temperature.

\begin{figure}
    \centering
    \includegraphics[width=1\linewidth]{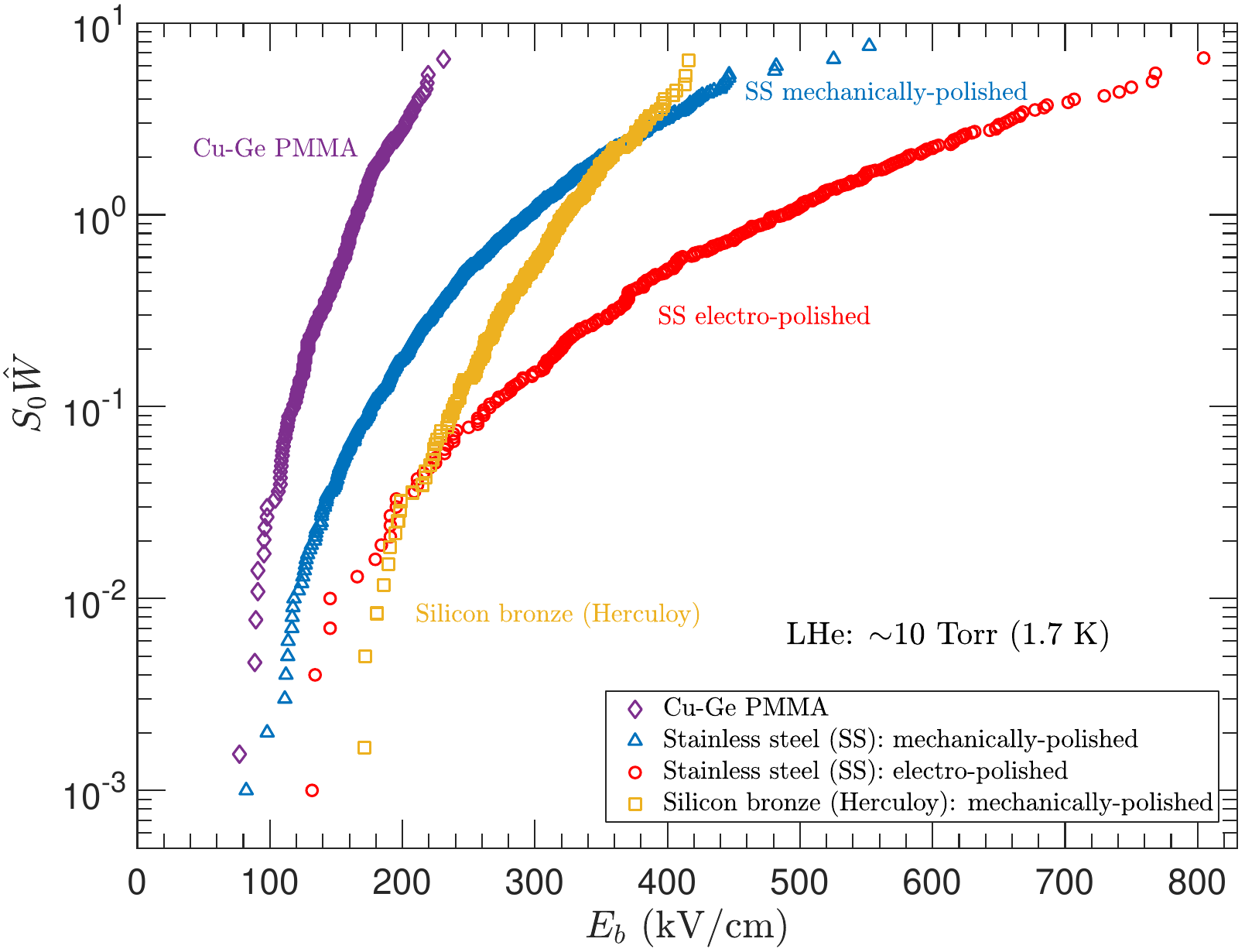}
    \caption{Breakdown hazard function measured for Cu-Ge coated PMMA, silicon bronze, and mechanically polished and electropolished stainless steel}
    \label{fig:SW_Cu-Ge-PMMA_Silicon_Bronze}
\end{figure}

Of these samples, the hazard function for electrical breakdown was measured for Cu-Ge coated PMMA and silicon bronze. The results are plotted in Fig.~\ref{fig:SW_Cu-Ge-PMMA_Silicon_Bronze}. Also plotted are the hazard function for mechanically polished and electropolished stainless steel. As seen in the figure, in the low electric field region---the region relevant to evaluating the expected performance of the electrode system (see Sec.~\ref{sec:expected_performance})---the hazard functions of Cu-Ge coated PMMA and silicon bronze fall in the range spanned by mechanically polished and electropolished stainless steel. 

\section{Expected performance and discussion}
\label{sec:expected_performance}
\begin{figure}
    \centering
    \includegraphics[width=1\linewidth]{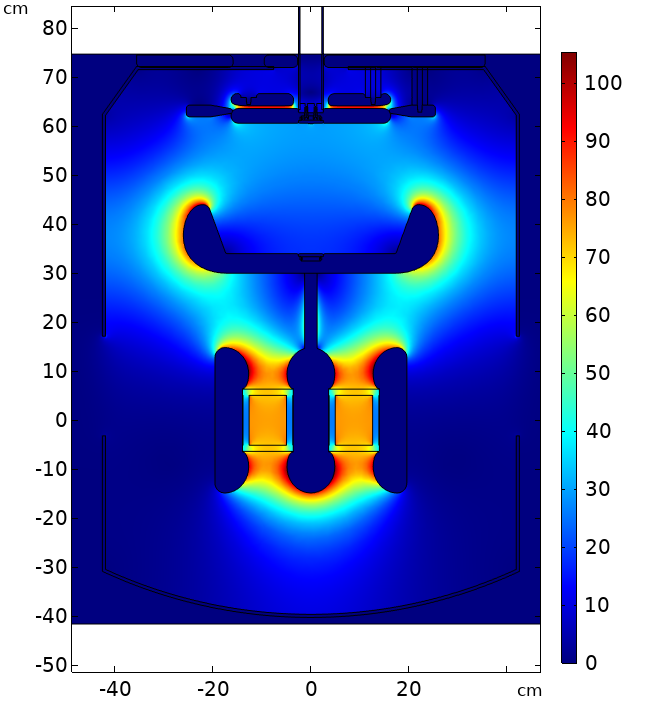}
    \caption{Electric field distribution in the Cryogenic nEDM apparatus.}
    \label{fig:nEDM_field_distribution}
\end{figure}

\begin{figure}
    \centering
    \includegraphics[width=1\linewidth]{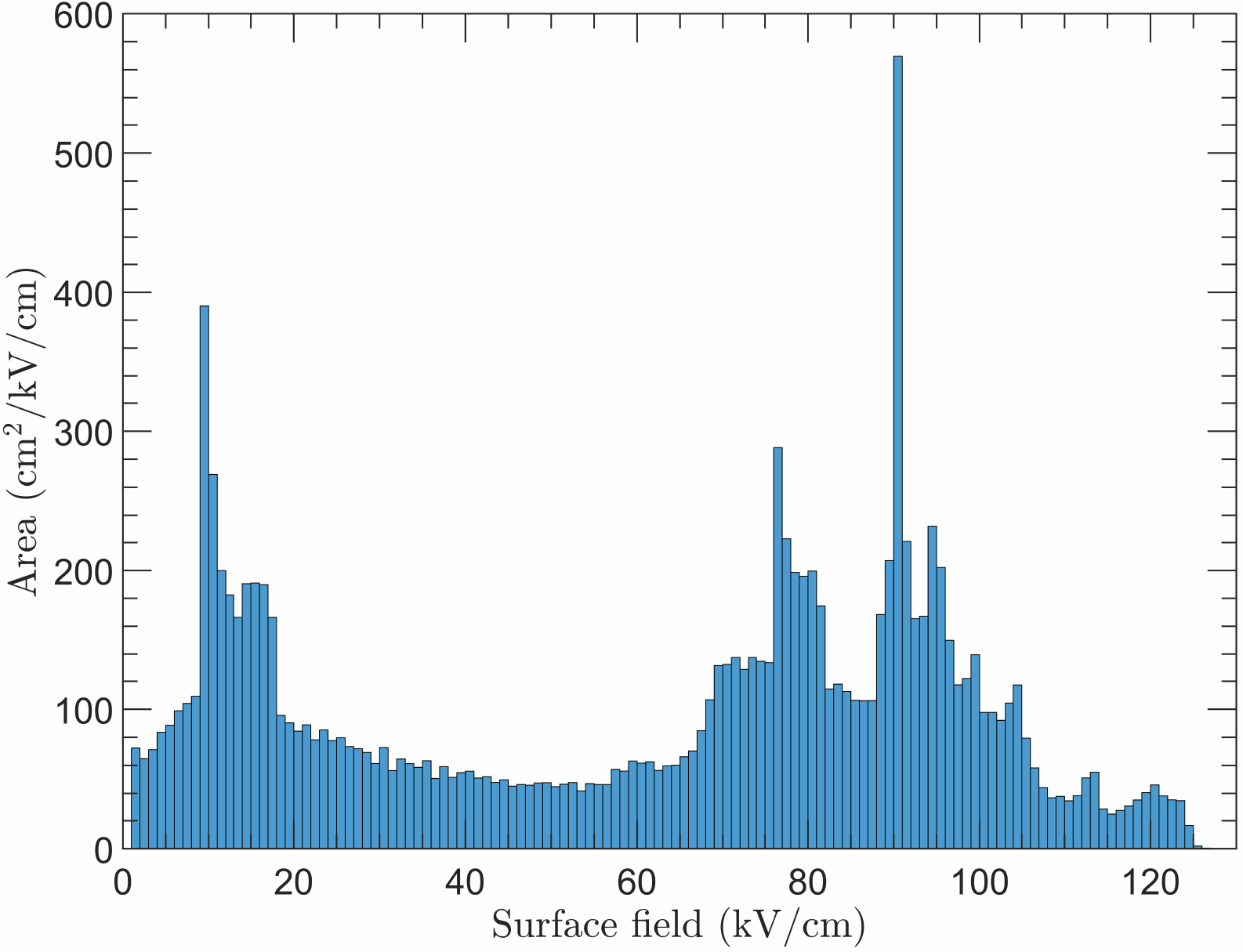}
    \caption{Surface area vs electric field for the electrode system of the Cryogenic EDM experiment.}
    \label{fig:nEDM_field_vs_area}
\end{figure}

Now using Eq.~(\ref{eq:scaledcdf_gen}) we can calculate the breakdown probability for the Cryogenic nEDM apparatus. $S(E)$, the electrode surface area at electric field $E$, is obtained by performing electrostatic FEM calculations using COMSOL. The electric field distribution is shown in Fig.~\ref{fig:nEDM_field_distribution}, and the area vs electric field is shown in Fig.~\ref{fig:nEDM_field_vs_area}. The measured hazard function is shown in Fig.~\ref{fig:SW_Cu-Ge-PMMA_Silicon_Bronze}.

\begin{figure}
    \centering
    \includegraphics[width=1.0\linewidth]{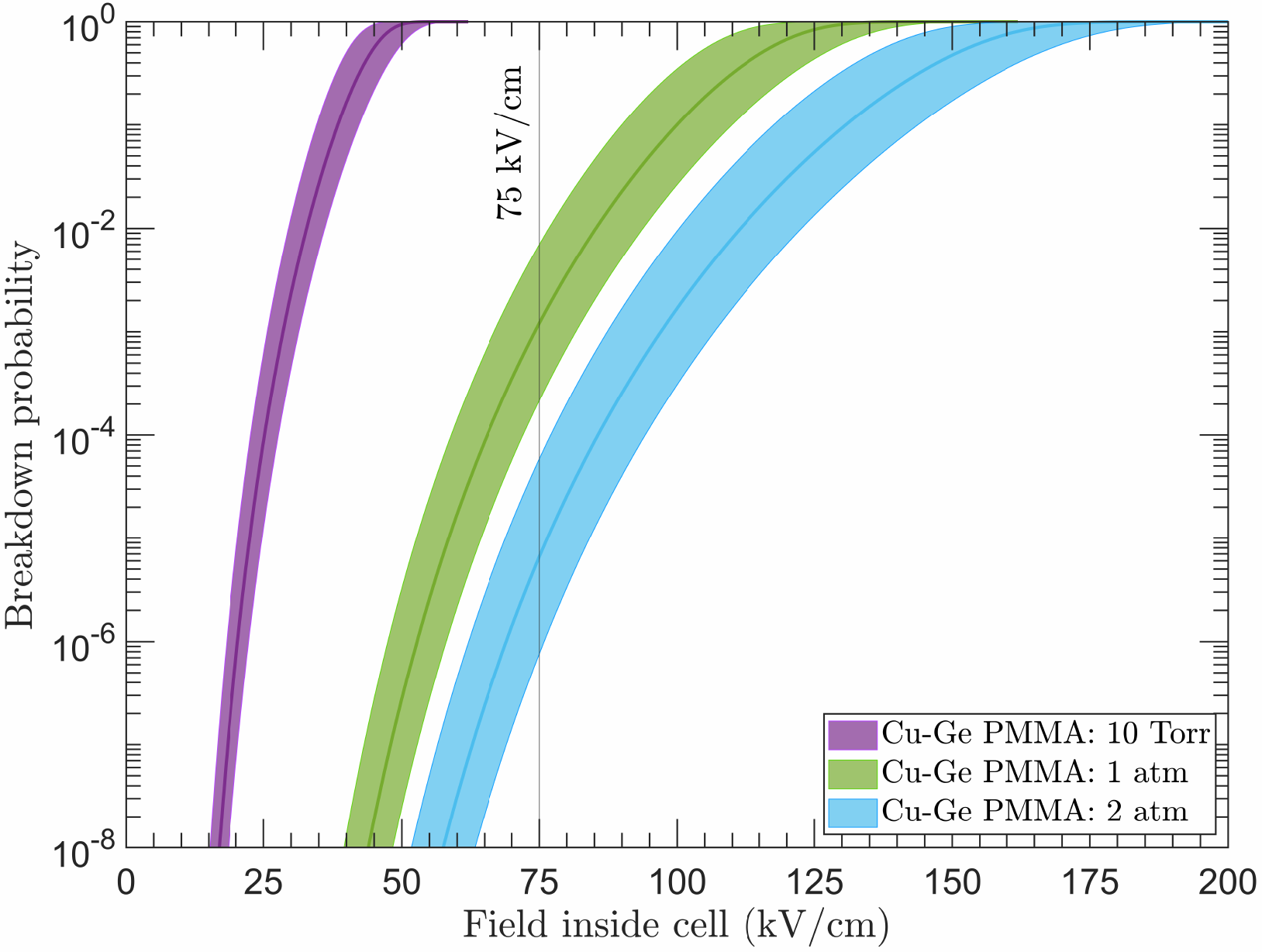}
    \caption{Breakdown probability calculated for Cu-Ge coated PMMA electrodes based on the measured hazard function.}
    \label{fig:PbCuGe}
\end{figure}

\begin{figure}
    \centering
    \includegraphics[width=1.0\linewidth]{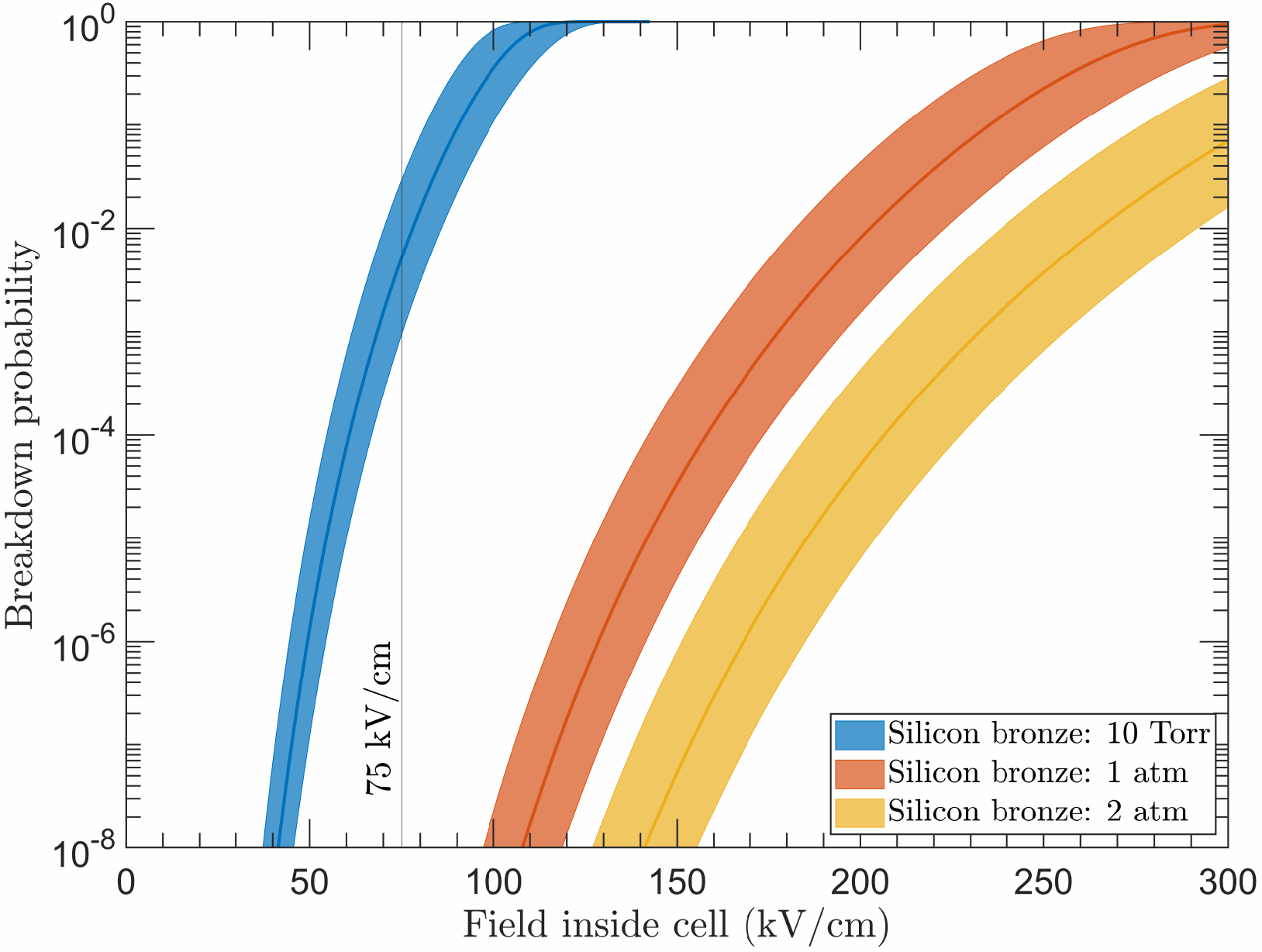}
    \caption{Breakdown probability calculated for silicon bronze electrodes based on the measured hazard function.}
    \label{fig:PbSiCu}
\end{figure}

The breakdown probabilities for the Cryogenic nEDM apparatus are shown in Figs.~\ref{fig:PbCuGe} and \ref{fig:PbSiCu}. Figure~\ref{fig:PbCuGe} corresponds to the case in which the electrodes are made of Cu-Ge coated PMMA, and Fig.~\ref{fig:PbSiCu} corresponds to the case in which the electrodes are made of silicon bronze. The uncertainty band for each curve reflects the 10\% uncertainty in the electric field determination in the small electrodes measurement of Ref.~\cite{Phan2021} (see Fig.~\ref{fig:sshv}~(a)) due to the uncertainty in the gap size at cryogenic temperatures. In both cases, the pressure dependence shown in Fig.~\ref{fig:sshv}~(d) is used to scale the hazard function measured at 10~torr to that for 2~atm. The pressure scaling has an effect of shifting the hazard function shown in Fig.~\ref{fig:SW_Cu-Ge-PMMA_Silicon_Bronze} to the right. The experiment adopted the operating pressure of 2~atm to reduce the breakdown probability. 

As seen in the figures, the central value of the breakdown probability for reaching 75~kV/cm is $6.6\times 10^{-6}$ and $< 10^{-16}$ for Cu-Ge coated PMMA electrodes and silicon bronze electrodes, respectively. Note that, as obvious from how the breakdown probability is defined in Eq.~(\ref{eq:cumulative_breakdown_distribution}) from the measured breakdown distribution, the breakdown probability discussed here refers to the probability of breakdown per voltage ramp. However, as discussed in Sec.~\ref{subsec:electrical_breakdown}, the breakdown probabilities of successive ramps are not necessarily statistically independent. That is, if a breakdown does not occur below a voltage $V_0$ during the first ramp, and the voltage is subsequently ramped down before breakdown occurs, then no breakdown occurs below $V_0$ during the second ramp. It is as if the condition of the electrode became ``reset'' every time a breakdown occurs but it remained the same if a breakdown does not occur even if the voltage is ramped up and down. After each reset, there is a new surface condition such that the next breakdown will follow the same statistical distribution~\footnote{Obviously this statement is only true if the stored energy in the system that gets released by the breakdown is not so large that it significantly alters the condition of the electrode surface}. The same observation was made in the experiment reported in Ref.~\cite{Ito2012}, which was performed to study scintillation of LHe in a high electric field. Therefore, the probability of observing a breakdown over the course of the experiment ($P_b^{\rm tot}$) is much smaller than what it would be if each successive ramp was independent, namely
\begin{equation}
    P_b^{\rm tot,\;ind} = 1- \left ( 1- P_b \right )^N,
\end{equation}
where $N$ is the number of ramps that occurs over the course of the experiment. In the Cryogenic nEDM experiment, about 10,000 measurement cycles are needed to achieve the stated goal sensitivity~\cite{Ahmed2019}. Therefore, $N$ is at most 10,000, which corresponds to a case the voltage is ramped for each measurement cycle. On the other hand, if the probability breakdown shown in Figs.~\ref{fig:PbCuGe} and \ref{fig:PbSiCu} applies only once, then we will have $P_b^{\rm tot} = P_b$. In reality, $P_b^{\rm tot}$ is likely to be somewhere in between but is much closer to $P_b$ than to $1-(1-P_b)^N$. That is
\begin{equation}
    P_b < P_b^{tot} \ll 1- \left ( 1- P_b \right )^N.
\end{equation}
We do not know the exact condition under which the electrode gets ``reset''. For example, it may get reset every time the apparatus is warmed up and cooled down. Nevertheless, we expect most of the ramps to be statistically dependent. 
Numerically,
\begin{equation}
\label{eq:total_breakdown_prob}
    6\times10^{-5} < P_b^{\rm tot} \ll 0.45\;\;{\rm (for\;Cu\textendash Ge\;coated\;PMMA)},
\end{equation}
for $N=10,000.$
For silicon bronze electrodes, $P_b < 10^{-16}$ and $1-(1-P_b)^N <1.1\times 10^{-12}$ for $N=10,000$. Again we emphasize that $P_b^{\rm tot}$ is expected to be much closer to the left hand side than to the right hand side of Eq.~(\ref{eq:total_breakdown_prob}). Therefore both materials give a total breakdown probability $P_b^{tot}$ acceptable for the experiment. 

It should be noted that when evaluating Eq.~(\ref{eq:scaledcdf_gen}) with the field distribution shown in Fig.~\ref{fig:nEDM_field_vs_area} to obtain Figs.~\ref{fig:PbCuGe} and \ref{fig:PbSiCu}, the highest electric field value is $\sim$120~kV/cm and the bulk of the contribution to Eq.~(\ref{eq:scaledcdf_gen}) comes from $E\sim 80-100$~kV/cm, which is at the low field end of the measured breakdown field distributions, especially after the hazard function in Fig.~\ref{fig:SW_Cu-Ge-PMMA_Silicon_Bronze} is shifted to the right to include the pressure effect. This explains the low probability of breakdown at $E_{\rm cell} =75$~kV/cm shown in Figs.~\ref{fig:PbCuGe} and \ref{fig:PbSiCu}, where $E_{\rm cell}$ refers to the electric field inside the measurement cells. 

As seen in Fig.~\ref{fig:Cavallo_250kV}, when the voltage on the HV electrode is ramped using the Cavallo multiplier, the voltage does not increase monotonically as it would when a power supply is used as was done to take the data shown in Fig.~\ref{fig:sshv}~(a). However, we expect that the breakdown probability given by Eq.~(\ref{eq:scaledcdf_gen}) applies because we did not observe breakdown probability dependence on the ramp rate, and because we did not observe breakdown while ramping down. 



Another important question to consider is how often breakdown events can occur while the system is held at the nominal operating voltage after it has been reached without breakdown. 
The Cryogenic EDM experiment requires maintaining a stable electric field for a minimum duration of 2000~s, which represents the combined time required for neutron filling and measurement. This requirement reflects an experimental operating procedure where the voltage is ramped for each measurement cycle.
However, the data presently available to fully address this question are somewhat limited.
Using the apparatus described in Ref.~\cite{Ito2016}, we maintained an electric field of $\sim$100~kV/cm in a 1-cm gap between two stainless steel electrodes with a cathode stressed area of 76~cm$^2$, immersed in LHe at $\sim$0.4~K at pressures between the SVP and $\sim$700~torr, for periods of 1000~s on several occasions. We observed occasional breakdowns, which appeared to originate in the HV feed system rather than in the gap between the electrodes. We also reported in Ref.~\cite{Phan2021} a measurement in which we held the system at a constant voltage for a predetermined amount of time (2~min) once the target voltage had been reached without breakdown. This procedure was repeated several hundred times. Except for cases in which the preset voltage was just below the voltage at which a breakdown would have been observed had the ramp continued---inferred from the comparison of the breakdown voltage distribution with and without voltage holding---no breakdown was observed during the holding period.  

It is evident that the current data does not allow for statistically significant inferences regarding the long-term stability of the electrode system for the Cryogenic EDM experiment. Due to the rare occurrence of these breakdown events, obtaining robust statistics is quite challenging.  In the absence of sufficient data,  we, instead, provide a qualitative discussion of phenomena frequently hypothesized to influence the long-term stability of high-voltage systems.

We first note that the HV and electrode system for the Cryogenic nEDM experiment is expected to operate at the lower tail end of the breakdown voltage distribution (see, e.g., Fig.~\ref{fig:SW_Cu-Ge-PMMA_Silicon_Bronze}). It follows that the mechanisms responsible for breakdown events that may occur while the system is held at the operation voltage are likely to differ from those that cause breakdowns during the voltage ramp; if the system successfully reaches an operating voltage that lies in the lower tail end of the breakdown voltage distribution, this implies that the electrode surfaces do not contain  asperities at which the electric field is sufficiently high to initiate a breakdown, and this condition is unlikely to change over time. Therefore, the breakdowns that occur while the voltage is held at the operational voltage are caused by extrinsic factors, which depend on specific details of the system. These factors may include cosmic rays and other ionizing radiation, vapor bubbles generated by heat input, suspended solid impurities, and electrical charging of dielectric surfaces within the system. We have already excluded the possibility of cosmic rays and other ionizing radiation as possible causes of breakdown in Sec.~\ref{subsec:cosmic_rays} Below we will discuss the remaining factors.  

As discussed in Ref.~\cite{Ito2016}, vapor bubbles created on the electrode surfaces due to residual heat in the electrodes or heat conducted through the HV feed system can trigger breakdowns. LHe is particularly susceptible to this effect compared to other liquid dielectrics because of its small heat of vaporization. In general, this effect can be mitigated through proper apparatus design and operation---specifically, by ensuring adequate thermal anchoring of the HV feed system and allowing sufficient cooling of the system before applying HV. 
In He-II, the exceptionally high thermal conductivity in combination with the operating pressure of 2 atm suppresses bubble formation, thus mitigating the probability of electrical breakdowns initiated by this particular mechanism.
Moreover, the use of the Cavallo multiplier eliminates the direct thermal connection between HV electrodes and the HV feed system, further suppressing this effect. 

Suspended solid impurities in a dielectric liquid can promote electrical breakdowns (see, e.g., Refs.~\cite{Gallagher1975,Kuffel1983}). 
These particles can become polarized when subjected to an electric field; over time, their accumulation on electrode surfaces leads to the formation of microprotrusions, which may serve as initiation sites for electrical breakdown.
In liquid $^4$He, all elements except $^3$He exist in solid form, so such contamination can occur readily. This effect can be mitigated by using chemically pure LHe.

Breakdowns caused by charge accumulation on dielectric surfaces have been reported in some systems~\cite{Abed_Abud_2022}. 
In the Cryogenic nEDM experiment, the measurement cells made of PMMA constitute the primary dielectric components. Ionizing radiation---primarily from neutron $\beta$ decay in the cold neutron beam during the UCN fill and from stored UCNs---can generate electric charge that can accumulate on the inner and outer surfaces of the cell walls. Charge produced inside the cells accumulates on the inner surfaces adjacent to the electrodes. Because the electric field inside the measurement cells is uniform, the charge accumulation is also uniform, resulting in a reduction of the electric field strength. Outside the measurement cells, charge generated from the neutron beam can accumulate on the outer surface of the walls oriented perpendicular to the electrodes, which can potentially cause HV instability. However, the collaboration has studied charge accumulation on PMMA surfaces in LHe and found that accumulated charge can be removed by simply changing the polarity of the electric field~\cite{Korsch2024}. 
Since the nominal operational mode of the experiment already requires regular polarity changes, we expect this effect not to present any significant risk to HV stability. 

Additional effects may contribute to HV instability, and their investigation will ultimately require a full-scale system. The analysis presented here justifies such an undertaking. 

Following the discussion on breakdown probability, it is also important to consider the possible impact of a breakdown. The capacitance of the electrode system is about 70~pF. Therefore the energy stored in the electrode system is 14~J at the electric potential of 635~kV. This significant energy is released if a breakdown occurs. Depending on how the energy is released and how quickly the energy is released, the breakdown can have various negative effects on the experiment. If all of 14~J goes to raising the temperature of 1500 l of LHe, the temperature of the LHe will increase from 0.4~K to 0.95~K. On the other hand, if it goes to raising the temperature of a small region of the silicon bronze electrode, it can vaporize approximately 3~mg of silicon bronze.  Depending on how quickly the energy is dissipated, the electrode surface could be severely damaged, and the transient emission associated with the breakdown could also upset electronics and other equipment used for the experiment. 

\begin{figure}
    \centering
    \includegraphics[width=\linewidth]{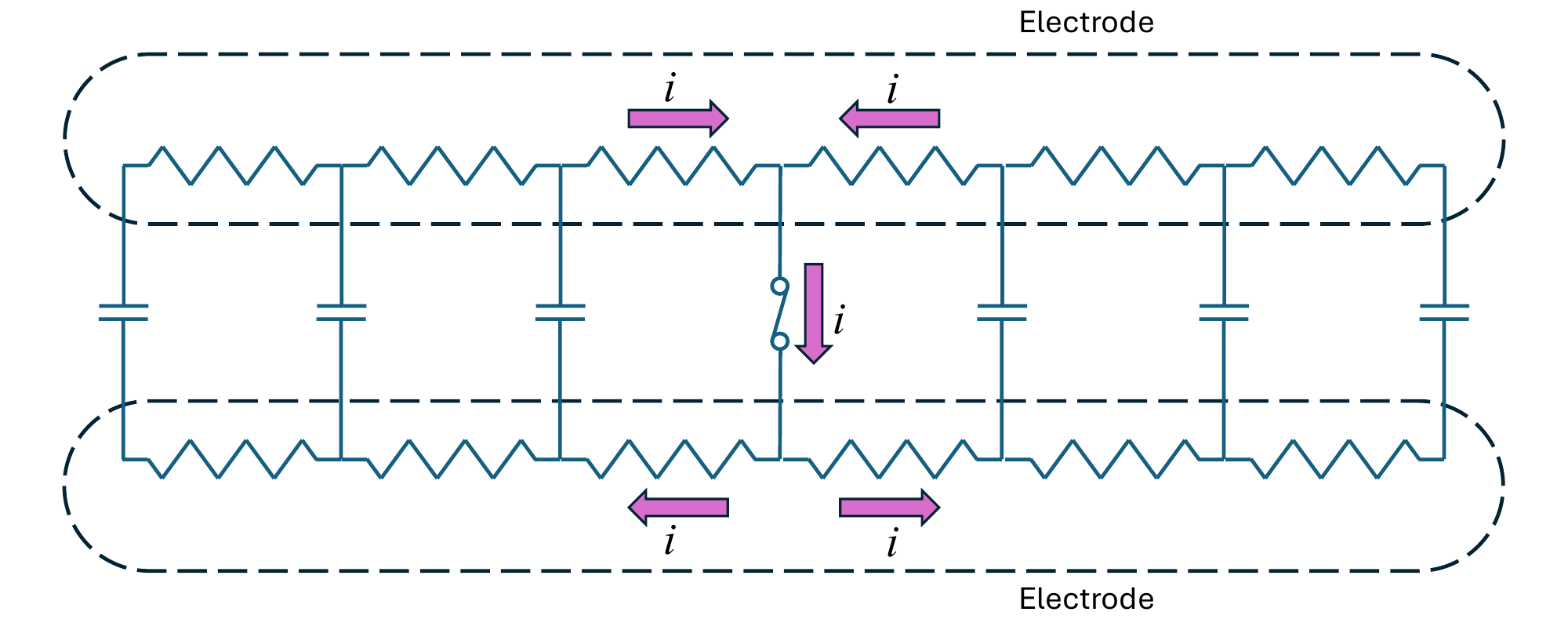}
    \caption{When the electrodes are made of resistive materials, the electrode system forms a distributed-parameter circuit in which resistors and capacitors are distributed. The figure shows an equivalent circuit. The electrical breakdown is represented by a closed switch. }
    \label{fig:breakdown_mitigation}
\end{figure}

The severity of the effect can be mitigated if the electrodes are made of resistive materials; the energy can be released in the form of Joule heating as electric currents flow through the resistive electrodes, while at the same time the resistive nature can slow down the process via $\tau = RC$, where $R$ is the effective resistance of the system (which depends on where the breakdown occurs) and $C$ is the capacitance of the system. In this regard, materials with even higher resistivities than those discussed in this paper may be preferable. Furthermore, the voltage drop associated with the current flowing through the resistive material will reduce the voltage at the breakdown site, potentially causing the breakdown to cease (a property that could be termed ``self-extinguishing''). These mitigating mechanisms are depicted in Fig.~\ref{fig:breakdown_mitigation}. From the robustness point of view, bulk electrode materials such as silicon bronze and silicon carbide are preferable to electrodes based on thin coatings such as Cu-Ge coated PMMA. 



\section{Conclusion and outlook}
\label{sec:discussion}
%
%
Through our R\&D program, we developed an understanding of several aspects of electrical breakdown in LHe that are relevant to the development of HV and electrode system for the Cryogenic nEDM experiment. These include: (1)~establishing that the breakdown in LHe is initiated on the electrode surface and can be suppressed by pressurization, and (2)~establishing the area scaling law for breakdown fields. Building on these findings, we developed a method for identifying and characterizing electrode materials as well as developing electrodes. Using this method, we designed an electrode system that satisfies the requirements of the Cryogenic nEDM experiment. We identified suitable electrode materials, namely Cu-Ge coated PMMA, silicon bronze, and silicon carbide. We also identified Cavallo's multiplier as a means to generate the required electric potential {\it in situ} within the LHe volume, and we experimentally demonstrated its functionality by constructing and operating a full-scale prototype. 

To move forward, a few remaining R\&D items must be completed before the full-scale electrode system can be constructed. The resistivity of the candidate materials at 0.4~K should be measured. Once confirmed, full scale electrodes based on Cu-Ge coated PMMA and on silicon bronze can be fabricated. In the case of SiC, a breakdown distribution measurement using small electrodes should be performed to verify adequate HV breakdown characteristics before full scale electrode construction begins. 

The materials and methods presented in this paper are also relevant to cryogenic nEDM experiments that do not employ the Golub-Lamoreaux method for detection of the nEDM signal. For example, the Ramsey separated oscillatory field technique, as used in the CryoEDM~\cite{Baker2010}, could be applied. The Cavallo multiplier is readily applicable to such an experimental approach. Likewise, the method we developed for identifying and characterizing electrode materials and for developing electrodes is also directly transferable. Furthermore, our method for calculating the magnetic Johnson noise for arbitrary geometries~\cite{Phan2024} will become important for cryogenic nEDM experiments, even for those not employing SQUIDs. 

In conclusion, we have established a solid foundation for HV and electrode systems for cryogenic nEDM experiments and demonstrated that the technology required to realize it is now in hand, making this approach one of the most promising paths toward achieving a sensitivity level of $10^{-28}$~$e\cdot$cm, owing to the significantly higher electric fields made possible by our work.

\section{Acknowledgments}
This work was supported by the United States Department of Energy, Office of Science, Office of Nuclear Physics under Contract Nos. 89233218CNA000001 (proposal LANLEEDM) and DE-AC05-00OR22725, the Los Alamos National Laboratory Integrated Contract Order No. 4000129433 with Oak Ridge National Laboratory, and the National Science Foundation Grants NSF-2110898, NSF-1822515, and NSF-1812340—``Fundamental Studies in Nuclear Physics.'' The authors thank V. Cianciolo for his leadership of the nEDM@SNS project and for his contributions to the development of Cu–Ge–coated PMMA electrodes and the evaluation of neutron activation. T.~M.~I. thanks P.~C.~Rowson for insightful discussions on electrical breakdown and bulk electrode materials.

\bibliography{cryogenic_EDM_HV}

\end{document}